\documentclass[preprint]{aastex631}

\shorttitle{Antarctic TianMu Survey Project}
\shortauthors{Zhou et al.}

\begin{document}

   \title{
   Antarctic TianMu Staring Observation Project I: Overview and Implementation of the Prototype Telescope}

\author[0009-0004-1255-6716]{Dan Zhou}
\affiliation{Shanghai Astronomical Observatory, Chinese Academy of Sciences, 80 Nandan Road, Shanghai 200030, PR China}
\author[0000-0001-5245-0335]{Jing Zhong}
\affiliation{Shanghai Astronomical Observatory, Chinese Academy of Sciences, 80 Nandan Road, Shanghai 200030, PR China}
\author{Jianchun Shi}
\affiliation{Shanghai Astronomical Observatory, Chinese Academy of Sciences, 80 Nandan Road, Shanghai 200030, PR China}
\author{Zhenghong Tang}
\affiliation{Shanghai Astronomical Observatory, Chinese Academy of Sciences, 80 Nandan Road, Shanghai 200030, PR China}
\author{Shiyin Shen}
\affiliation{Shanghai Astronomical Observatory, Chinese Academy of Sciences, 80 Nandan Road, Shanghai 200030, PR China}
\author{Peng Jiang}
\affiliation{Key Laboratory for Polar Science, MNR, Polar Research Institute of China, Shanghai, 200136, China}
\author{Jie Zhu}
\affiliation{Shanghai Astronomical Observatory, Chinese Academy of Sciences, 80 Nandan Road, Shanghai 200030, PR China}
\author{Yong Yu}
\affiliation{Shanghai Astronomical Observatory, Chinese Academy of Sciences, 80 Nandan Road, Shanghai 200030, PR China}
\author{Lixin Zheng}
\affiliation{Shanghai Astronomical Observatory, Chinese Academy of Sciences, 80 Nandan Road, Shanghai 200030, PR China}
\author{Jianjun Cao}
\affiliation{Shanghai Astronomical Observatory, Chinese Academy of Sciences, 80 Nandan Road, Shanghai 200030, PR China}
\author{Guoping Chen}
\affiliation{Shanghai Astronomical Observatory, Chinese Academy of Sciences, 80 Nandan Road, Shanghai 200030, PR China}
\author{Xinyu Yao }
\affiliation{Shanghai Astronomical Observatory, Chinese Academy of Sciences, 80 Nandan Road, Shanghai 200030, PR China}
\author{Congcong Zhang}
\affiliation{Shanghai Astronomical Observatory, Chinese Academy of Sciences, 80 Nandan Road, Shanghai 200030, PR China}
\author{Lurun Shen}
\affiliation{Shanghai Astronomical Observatory, Chinese Academy of Sciences, 80 Nandan Road, Shanghai 200030, PR China}
\author{Hui Zhang}
\affiliation{Shanghai Astronomical Observatory, Chinese Academy of Sciences, 80 Nandan Road, Shanghai 200030, PR China}
\author{Xiang Pan}
\affiliation{Key Laboratory for Polar Science, MNR, Polar Research Institute of China, Shanghai, 200136, China}
\author{Chenwei Yang}
\affiliation{Key Laboratory for Polar Science, MNR, Polar Research Institute of China, Shanghai, 200136, China}
\author{Tuo Ji}
\affiliation{Key Laboratory for Polar Science, MNR, Polar Research Institute of China, Shanghai, 200136, China}
\author{Xian Shi}
\affiliation{Shanghai Astronomical Observatory, Chinese Academy of Sciences, 80 Nandan Road, Shanghai 200030, PR China}
\author{Hengxiao Guo}
\affiliation{Shanghai Astronomical Observatory, Chinese Academy of Sciences, 80 Nandan Road, Shanghai 200030, PR China}
\author{Zhen Yan }
\affiliation{Shanghai Astronomical Observatory, Chinese Academy of Sciences, 80 Nandan Road, Shanghai 200030, PR China}
\author{Donghai Zhao}
\affiliation{College of Physics and Electronic Information Engineering, Guilin University of Technology, Guilin 541004, China}
\affiliation{Key Laboratory of Low-dimensional Structural Physics and Application, Education Department of Guangxi Zhuang Autonomous Region, Guilin 541004, China}
\author{Liang Chen}
\affiliation{Shanghai Astronomical Observatory, Chinese Academy of Sciences, 80 Nandan Road, Shanghai 200030, PR China}
\author{Jianeng Zhou }
\affiliation{Shanghai Astronomical Observatory, Chinese Academy of Sciences, 80 Nandan Road, Shanghai 200030, PR China}
\author{ Minfeng Gu}
\affiliation{Shanghai Astronomical Observatory, Chinese Academy of Sciences, 80 Nandan Road, Shanghai 200030, PR China}
\author{Fuguo Xie}
\affiliation{Shanghai Astronomical Observatory, Chinese Academy of Sciences, 80 Nandan Road, Shanghai 200030, PR China}
\author{Wenbiao Han}
\affiliation{Shanghai Astronomical Observatory, Chinese Academy of Sciences, 80 Nandan Road, Shanghai 200030, PR China}
\author{Jinliang Hou}
\affiliation{Shanghai Astronomical Observatory, Chinese Academy of Sciences, 80 Nandan Road, Shanghai 200030, PR China}
\author{Bixuan Zha}
\affiliation{Shanghai Astronomical Observatory, Chinese Academy of Sciences, 80 Nandan Road, Shanghai 200030, PR China}
\author{Wenwen Zuo}
\affiliation{Shanghai Astronomical Observatory, Chinese Academy of Sciences, 80 Nandan Road, Shanghai 200030, PR China}
\author{Chun Xu}
\affiliation{Shanghai Astronomical Observatory, Chinese Academy of Sciences, 80 Nandan Road, Shanghai 200030, PR China}
\author{ Zhengyi Shao}
\affiliation{Shanghai Astronomical Observatory, Chinese Academy of Sciences, 80 Nandan Road, Shanghai 200030, PR China}
\author{Lei Hao}
\affiliation{Shanghai Astronomical Observatory, Chinese Academy of Sciences, 80 Nandan Road, Shanghai 200030, PR China}
\author{Jian Fu}
\affiliation{Shanghai Astronomical Observatory, Chinese Academy of Sciences, 80 Nandan Road, Shanghai 200030, PR China}



\begin{abstract}
Wide-field rapid sky surveys serve as critical observational methods for time-domain astronomical research. The Antarctic region, with several months of continuous dark nights annually, is an ideal site for time-domain astronomical observations. The Antarctic TianMu Staring Observation Project aims to deploy a fleet of small telescopes, adopting an array observation model to conduct time-domain optical observations in Antarctica, featuring wide-sky coverage, high-cadence sampling, long-period staring, and simultaneous multi-band measurements. Considering the severe challenges optical telescopes face in Antarctica, including extremely low temperatures, unattended operation, and limited power supply and network transmission, we have designed and developed the Antarctic TianMu prototype telescope based on drift-scan charge-coupled device technology. In October 2022, our prototype (with an aperture of 18~cm), named AT-Proto was transported to Zhongshan Station in Antarctica aboard China’s 39th Antarctic Research Expedition. It has since operated stably and reliably in the frigid environment for over two years, demonstrating the significant advantages of this technology in polar astronomical observations. The experimental observation results of AT-Proto provide a solid foundation for the subsequent construction of a time-domain astronomy observation array in Antarctica.

\end{abstract}



\section{Introduction} 
\label{sect:intro}

Astronomy is a fundamental discipline driven by observational measurements. With technological advancements, contemporary astronomical observations have evolved from characterizing the static universe to understanding its dynamic nature. Revealing the variability of celestial objects and discovering new astronomical phenomena through multi-band, continuous/rapid observations defines time-domain astronomy -- an emerging astronomical subdiscipline \citep{2025RAA....25d4009H}. This field has yielded a plethora of significant scientific discoveries, including observations of supernovae, flare stars, active galactic nuclei (AGN), gravitational wave events, exoplanets, compact objects, and other key targets. It is foreseeable that within the next 10 to 20 years, time-domain astronomy will become a leading, frontier in international astronomy, continuing to provide a wealth of information \cite{2002SPIE.4836...10T,2010ApJ...713L..79K}.

Among numerous sky survey projects, the Zwicky Transient Facility (ZTF) and the Large Synoptic Survey Telescope (LSST) stand out, each with distinctive characteristics in their sky survey sampling strategies. The ZTF uses the Palomar 48-inch Schmidt telescope. Endowed with an ultra-large field of view of 47 square degrees, enabled by a 600-megapixel detector, it can scan the entire northern hemisphere sky once every three observing nights, allowing for the rapid detection of a large number of transient events \citep{2019PASP..131a8002B,2019PASP..131f8003B}.  The LSST uses an 8.4-meter primary mirror and boasts a field of view of up to 9.6 square degrees. Its limiting magnitude for a single observation reaches 24th magnitude in a 10-second exposure, and it can survey up to 14,000 square degrees of the sky once every three days \citep{2002SPIE.4836...10T}. This makes the LSST capable of observing much fainter celestial objects. It is expected to yield significant scientific outcomes across multiple fields, and its abundant data will provide greater information and possibilities for astronomical research \citep{2008SerAJ.176....1I}.

China has also made remarkable achievements in the field of time-domain sky surveys. The Wide Field Survey Telescope (WFST) is a 2.5-meter aperture time-domain sky survey telescope with a 3-degree-diameter field of view (FOV). During its commissioning and pilot observation phases, it provided minute-cadence observations (with a total on-source time of approximately 13 hours for specific targets) and surveyed two-thirds of the Northern Hemisphere every three days, boasting capabilities of wide-field, high-precision, and wide-wavelength sky surveys \citep{2023SCPMA..6609512W,2024RAA....24a5003C,2025ApJS..278...29L}. As the world’s first wide-field multi-channel photometric sky survey telescope, the Mephisto Telescope has a FOV with a diameter of 2${\deg}$. It can capture celestial images simultaneously in three channels, enabling it to provide ultra-high-precision photometry and color information of celestial objects. This facilitates the real-time acquisition of celestial color data, enhancing the reliability and efficiency of identifying and classifying variable and transient sources \citep{2019gage.confE..14L, 2022RAA....22b5004L}.

Time-domain astronomical observations are currently advancing in three key directions \cite{2024FrASS..1104616H}. The first is shorter timescales, which require near-continuous monitoring; the second is larger fields of view, demanding coverage of the widest possible celestial area, and; the third is enhanced detection capabilities, enabling observation of fainter celestial objects.
However, existing time-domain sky surveys have two notable limitations. First, the quantity and geographic distribution of observation equipment are suboptimal. This makes continuous monitoring impossible. The other limitation is that the fastest current survey cycles still take at least one day, which severely hinders the monitoring of astronomical phenomena which vary on shorter timescales \cite{2019PASP..131a8002B,2016arXiv161205560C,2002SPIE.4836...10T,2023SCPMA..6609512W,2020SPIE11445E..7MY}. 

Transient sources with extremely short timescales (on the order of milliseconds to hours) are central to contemporary astronomical research. Sub-daily phenomena include fast radio bursts \cite{2019A&ARv..27....4P}, gravitational wave sources \citep{2016PhRvL.116f1102A}, supernova shock breakouts \cite{2010ApJ...716..781K,2012ApJ...747...88N}, gamma-ray bursts \cite{2004RvMP...76.1143P}, supernovae \cite{2013ApJ...778..164A}, tidal disruption events of stars by massive black holes \cite{2015JHEAp...7..148K}, extremely bright X-ray bursts \cite{2008Natur.453..469S}, superflare stars \cite{2013ApJS..209....5S,2019ApJS..241...29Y}, and classical novae \cite{2008clno.book.....B}. All of these are active areas of international research in optical astronomy. To unravel the origins and physical mechanisms of these phenomena, astronomers urgently need observational data on shorter timescales. Reducing survey cycles to hours or less would greatly facilitate the real-time capture of transient events and acquisition of critical data at their onset.

A possible approach to address the first limitation is to deploy telescopes in Antarctica,  which has been demonstrated to be the best site on Earth for ground-based optical observations \citep{2020RAA....20..168S}. Due to its unique geographical location and environment, Antarctica has several distinctive advantages for short-timescale time-domain astronomical observations. It is a promising location because it experiences up to hundreds of days of polar night each year, allowing 24-hour daily observations, which facilitates capturing the earliest stages of rapid celestial changes and transient sources in real time. Many Antarctic sites also have a high number of clear nights \textcolor{red}{\citep{2021MNRAS.501.3614Y}}, enabling months-long uninterrupted observations that other locations cannot achieve. In addition, the Antarctic night sky background is extremely dark \textcolor{red}{\citep{2010AJ....140..602Z}}. Except for scientific expeditions, Antarctica is uninhabited, with no light pollution at night, making it highly suitable for optical astronomical observations. Finally, the atmosphere at Antarctic sites has high transparency, because Antarctica is free from air pollution or natural phenomena like sandstorms. These factors combined make it an ideal location for optical astronomy \cite{2008SPIE.7012E..4GY,2014SPIE.9145E..0FY,2018SPIE10700E..1LL,2020Natur.583..771M}. 

It is worth noting that, in addition to site testing work, a substantial number of pioneering astronomical observation projects have been carried out in Antarctica, achieving a series of important results and accumulating valuable experience for subsequent Antarctic astronomical observations. Among the projects in Antarctic astronomy, the Chinese Antarctic Center Telescope Array (CSTAR) leads in continuous astronomical observations in the Antarctic interior, demonstrating the operational stability of equipment in extreme environments \citep{2008SPIE.7012E..4GY, 2010PASP..122..347Z}. The Antarctic Survey Telescope (AST3), relying on its wide field of view and high-sensitivity observational capabilities, has made important breakthroughs in the field of time-domain astronomy, such as participating in the early detection of supernovae and detection of the optical counterparts to gravitational wave events  \citep{2008SPIE.7012E..2DC, 2017SciBu..62.1433H,2018MNRAS.479..111M,2023MNRAS.520.5635Y}.

A possible approach to address high-cadence sampling is to develop multiple small-aperture wide-field telescopes \citep{2025RAA....25d4001H,2021AnABC..93..628L}. By forming an observation array, the field of view (FoV) can be matched to the survey area, eliminating the need for frequent changes in the telescope's exposure region. This enables staring observations and short-timescale repeated sampling of the surveyed sky area. Additionally, dual-color synchronous photometric surveys can be achieved by simultaneously deploying two sets of telescope arrays for red and blue bands at the same site.

In this project, by constructing the Antarctic TianMu Staring Observation Platform composed of 30 small-aperture wide-field telescope arrays in Antarctica, we will carry out minute-scale sampling and dual-color synchronous observations for hundreds of consecutive days. The targets will be celestial objects brighter than magnitude 18 (5-minute exposure time) in a 1,200-square-degree sky area around the South Celestial Pole ($ -75{\deg} < \delta < -65 {\deg}$). The construction of this observation platform will effectively fill the gap in short-timescale time-domain astronomical observations. This initiative utilizes the unique advantages of Antarctica, with long polar nights, dark skies, and high atmospheric transparency, to enable continuous, high-cadence monitoring of cosmic transients and time-domain objects, helping to elucidate the dynamic universe.

\section{Scientific Objectives}

The Antarctic TianMu Staring Observation Project (ATSOP) performs continuous polar-night observations, combined with high-cadence sampling, to conduct statistical studies on galactic time-domain objects, high-energy transient sources, and high-inclination small bodies in the solar system. By acquiring comprehensive time-series observation data of various time-domain objects, the aim is to fill the gap in time-domain astronomical observation modes and deepen observational research in stellar physics, solar system origin, and fundamental astronomy. In the next sub-sections, we introduce these four primary scientific objectives one by one. 

\subsection{Galactic Time-Domain Objects}
There are various types of galactic time-domain object, including compact object systems such as low-mass X-ray binaries and cataclysmic variables \cite{2011ApJ...729....8Z,2003cvs..book.....W}; variable and pulsating stars like Cepheid variables, RR Lyrae stars and flare stars \citep{1958HDP....51..353L,2020ApJS..249...18C}; binary systems such as eclipsing binaries and close binaries \cite{2002MNRAS.329..897H,2010ApJS..190....1R}, as well as red giants suitable for asteroseismology research \cite{2017EPJWC.16004003Z,2015AJ....149...84Z}, and; late-type stars with photometric variations due to magnetic activities\cite{2014ApJ...797..121H}. Their brightness or radiation characteristics change over time, containing rich astrophysical information.

The Antarctic TianMu Telescope experiences prolonged periods of uninterrupted polar-night, allowing continuous observation and minute-scale high-cadence sampling, providing unique advantages in Galactic time-domain astrophysics research. The polar night environment eliminates day-night interference, enabling long-term sequential tracking of target objects, allowing detailed observations of the complete evolutionary process of low-mass X-ray binaries or cataclysmic variables, from outbursts to quiescent states \cite{2018MNRAS.479.2777R}, as well as the oscillation signals of late-type stars in asteroseismology \citep{2010ApJ...713L.169C}. Minute-level high-cadence sampling can also accurately resolve the periodic oscillations of matter inflow in the inner region of accretion disks, the details of eclipse phases in binary systems \cite{2013EAS....64..269B,2002A&A...391..213K}, and even the minute-level luminosity changes of superflares \cite{2025arXiv250620540O}. In addition, the combination of large sky coverage and high-cadence sampling allows for batch monitoring of thousands of variable stars (such as Cepheid variables and RR Lyrae stars) in the southern sky. By constructing high-precision period-luminosity relationships through continuous light curves, its data integrity and time resolution far exceed those of traditional ground-based sky surveys. This unique combination of polar night continuous observation with high-cadence sampling enables the ATSOP to fill the gaps in current time-domain observations in terms of short-time-scale photometric variations and long-term sequential coverage. This capability furnishes pivotal empirical data for deciphering accretion dynamics in compact object systems, characterizing stellar interior architectures, and advancing variable star astrophysics \cite{2021AnABC..93..628L}.

\subsection{High-Energy Transient Sources}
There are numerous extremely high-energy transient source phenomena in the universe. Although various time-domain sky survey projects have been carried out internationally, the limitations of survey sampling patterns and ground-based observation durations mean that a large volume of time-domain observation data lacks early light curve information \cite{2020MNRAS.497.1925G,2013A&A...557A..12Z}. These data are precisely the key to studying such transient celestial phenomena and distinguishing different physical models \cite{2023PASP..135c4102K}. Therefore, time-domain observations of short-timescale transient sources are essential for exploring new celestial bodies and revealing new physics.

Using the ATSOP, we will be able to use high-cadence sampling (minute-level)  to conduct continuous time-domain sky surveys during polar night, over a large sky area of approximately 1,200 square degrees. We expect to obtain batches of early-time domain data for such extremely high-energy transient sources, including gravitational wave counterparts, supernovae, gamma-ray bursts, black hole tidal disruption of stars, ultraluminous X-ray bursts, and fast radio bursts, as well as potentially finding other unknown short-timescale transient source phenomena. This will provide key data support for further studying the origin mechanisms and evolutionary laws of various extremely high-energy astrophysical phenomena. In addition, a uniform and continuous short-timescale time-domain sky survey database will also provide a complete observational sample for us to further estimate the occurrence probabilities of various transient source objects. This is of great significance for distinguishing between different physical origin models \cite{2008Natur.455..183R}.

\subsection{High-Inclination Small Bodies In The Solar System}
Most small bodies in the Solar System are located near the ecliptic plane. Among known objects, those with high orbital inclination (i $>$ 30${\deg}$) account for less than 1\%. High-inclination small bodies include Near-Earth Objects (NEOs), Main Belt asteroids, Centaurs, Trans-Neptunian objects, long-period comets, and interstellar small bodies \cite{2012A&ARv..20...56C}. Limited by the lack of samples and observational data, there are few studies on the physical properties, origin and evolution of high-inclination small bodies \cite{2020MNRAS.494.2191N}. Carrying out observational research can improve understanding of their surface characteristics, origin and evolution, and the activity characteristics and evolutionary laws of active small bodies. This can provide a more comprehensive understanding of the properties and behaviors of small bodies to support scientific research, astronomical observation, space exploration, and planetary defense.

Because of the unique geographical location and the characteristics of a large field of view and high-cadence sampling, the Antarctic TianMu Telescope is suitable for observing and studying NEOs and other small solar system objects with high orbital inclination. Proposed work includes searching for and discovering new small bodies, obtaining physical properties such as the size, shape and rotation period of the observed small bodies, acquiring brightness variation and activity information of comets and active asteroids, and studying their activity.

\section{Observation Strategy}
\begin{figure*}[htbp]
    \centering
    \includegraphics[width=0.85\textwidth]{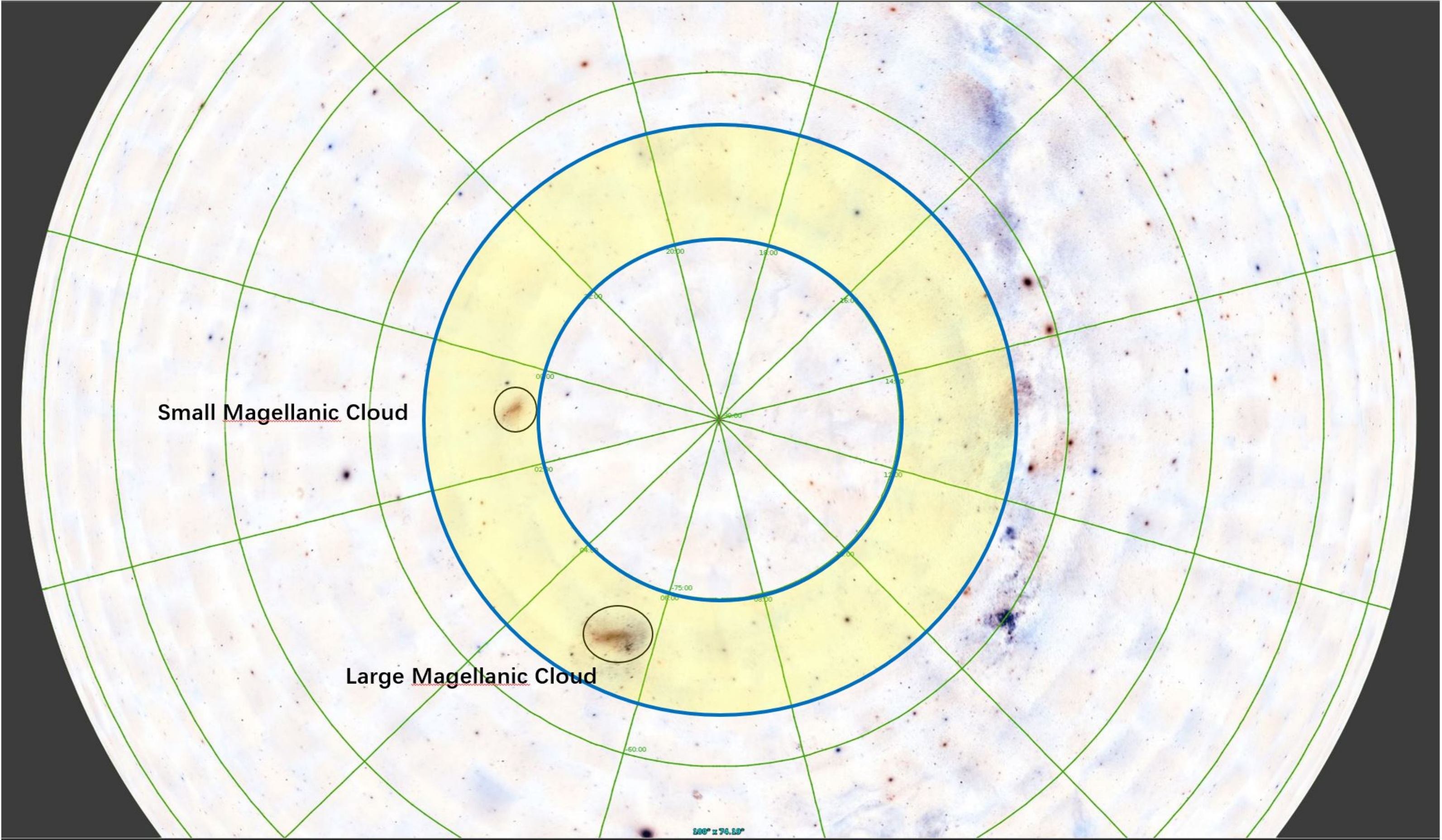}
    \caption{ Equatorial coordinate sky map near the South Celestial Pole. The staring area of ATSOP is marked in yellow, with a total area of 1,229 square degrees, encompassing part of the Galactic Disk, part of the high galactic latitude region, and the Large and Small Magellanic Clouds.}
    \label{fig:map}
\end{figure*}

The ATSOP observation region is defined as an annular area near the South Celestial Pole with declination ranging from -75${\deg}$ to -65${\deg}$, corresponding to galactic latitudes of -52.1${\deg}$ to -2.1${\deg}$ and ecliptic latitudes of -88.4${\deg}$ to -41.5${\deg}$. Figure~\ref{fig:map} shows the staring observation region of ATSOP. This region offers multiple advantages. The low galactic latitude area penetrates the Galactic disk, where dense stellar distributions provide abundant samples for monitoring photometric variations of time-domain objects such as variable stars and binary systems. Concurrently, the high galactic latitude halo region, with minimal interstellar medium interference, enables the capture of more extragalactic signals, significantly enhancing the probability of detecting high-energy transient sources like gamma-ray bursts and supernovae. Additionally, the high ecliptic latitude effectively avoids interference from the solar system's asteroid belt near the ecliptic plane, making it particularly suitable for tracking various solar system small bodies with high orbital inclinations. The region also completely covers the Large and Small Magellanic Clouds, allowing precise determination of galactic distances and exploration of dark matter distribution through continuous observation of variable stars, comprehensively supporting the research goals of time-domain astronomy and astrophysics.

The ATSOP observation region spans 10${\deg}$ in declination (-75${\deg}$ to -65${\deg}$), corresponding to a total sky area of 1,229 square degrees. To ensure the Antarctic TianMu Telescope array can observe the region at any time, we plan to deploy up to 15 Antarctic TianMu Telescopes for full coverage (each telescope has a field of view of 100 square degrees with a 20\% sky area repetition rate to enhance data reliability). To acquire real-time color data of celestial objects, the project plans to use two groups of Antarctic TianMu Telescopes (30 telescopes in total) with different band observation capabilities to conduct simultaneous parallel observations in the B (450~nm~--~650~nm) and R (650~nm~--~850~nm) optical bands for this sky area. This multi-band collaborative observation mode can not only accurately distinguish stellar types and track the evolution of variable stars through color indices (B-R), but also provide key photometric color change curves for high-energy transient sources, such as gamma-ray burst afterglows and supernova explosions. This will provide a multi-dimensional data foundation for time-domain astronomy research.

\begin{table*}[htbp]
    \centering
    \caption{Telescope Technical Specifications}
    \begin{tabular}{|l|l|}
        \hline
        \textbf{Parameter} & \textbf{Specification} \\
        \hline
        Telescope Aperture & 180~mm \\
        \hline
        Field of View & 9.5${\deg}$~×~9.5${\deg}$ \\
        \hline
        Focal Length & 220~mm \\
        \hline
        Bandpass & 500~--~800~nm \\
        \hline
        FWHM & Avg. 2 pixels; 11.25"/pixel \\
        \hline
        CCD & 12~$\mu$m/pixel; 3056~×~3056; 36.7~×~36.7~mm \\
        \hline
        QE @550~nm & 64\% \\
        \hline
        A/D Resolution & 16-bit \\
        \hline
        Read Noise & 14 e$^-$ \\
        \hline
        Dark Current @-20${\deg}$C & 0.27 e$^-$/pix/s \\
        \hline
        Full-well Capacity & 110~ke$^-$ \\
        \hline
        Drift Scan Speed Range & 1 - 1000~ms/line \\
        \hline
        Cooling System & Dual-stage TEC \\
        \hline
        Operating Temperature Range & -40${\deg}$C to +30${\deg}$C \\
        \hline
    \end{tabular}
    \label{tab:telescope-specs}
\end{table*}

\section{Technical Scheme of AT-Proto}

Conducting optical astronomical observations in Antarctica requires addressing five primary challenges of extreme low temperatures, high-wind conditions, remote unattended operation, limited power supply, and constrained network bandwidth \cite{2008SPIE.7012E..4GY}. These constraints must be integrally considered in the design of the telescope opto-mechanical system, receiver terminal, and remote-control architecture. For optical design, balancing a wide field of view (WFOV) with detection sensitivity necessitates image stacking, a well-established technique using multiple WFOV telescopes with small apertures (on the order of 180~mm). System design prioritizes minimizing moving mechanisms to reduce electronic components, thereby decreasing power consumption, enhancing reliability, and lowering costs of operation and maintenance.

Given the massive volume of raw image data collected by the receiver terminal, real-time preprocessing (e.g., background subtraction and source extraction) must occur locally. The processed results are significantly compressed, facilitating efficient data transfer for rapid alerts of transient events, such as supernovae or gravitational wave counterparts, to coordinate follow-up observations. Due to severe power constraints, the system prioritizes thermal energy recycling to maximize limited electricity usage.

Under these considerations, AT-Proto has been developed based on drift-scan Charge-Coupled Device (CCD) technology, with a core approach encompassing four key aspects. Drift-scan CCD cameras are installed on small-aperture wide-field telescopes, using the drift-scan technique to eliminate tracking mechanisms while enabling faint-object detection and ensuring high reliability. Concurrently, the entire optical telescope system is housed in a thermally regulated chamber, together with the control computers. This chamber is equipped with insulated windows and resistive heating elements, removing the need for cryogenic components and reducing construction costs. Local image preprocessing is also performed, so that only processed results transmitted, eliminating high-bandwidth dependencies. Finally, waste heat from CCD cameras and computers is utilized as auxiliary heating, achieving nearly 100\% energy efficiency with minimal power consumption.

AT-Proto consists of a temperature-controlled dome, a telescope system equipped with a drift-scan CCD camera, and a remote operation and control system. The telescope has a 180~mm aperture and is outfitted with a CCD camera containing a 3k~×~3k pixel array. Detailed specifications of the telescope system are provided in Table~\ref{tab:telescope-specs}. 

\subsection{Temperature-Controlled Dome}
The extreme polar cold can easily cause failures in telescope moving parts, cameras, or optical imaging systems, so we adopted an integrated design approach. The  telescope, camera, control system, and computers are housed within a temperature-controlled dome with insulated observation windows. The dome has a maximum outer diameter of 868~mm, and a height of 990~mm. It is placed on a 1m~×~1m base, which is equipped with support feet with a height of 700~mm, to allow wind and snow to pass without accumulating at the lower part of the dome. The dome is equipped with an optical window, which is made of fused silica and has an aperture of $\phi$260~mm. The telescope observes through this optical window, which is coated with ITO films on both surfaces. Heating the films raises the temperature, preventing frost and condensation in polar conditions.

\begin{figure}[htbp]
    \centering
    \includegraphics[width=0.5\textwidth]{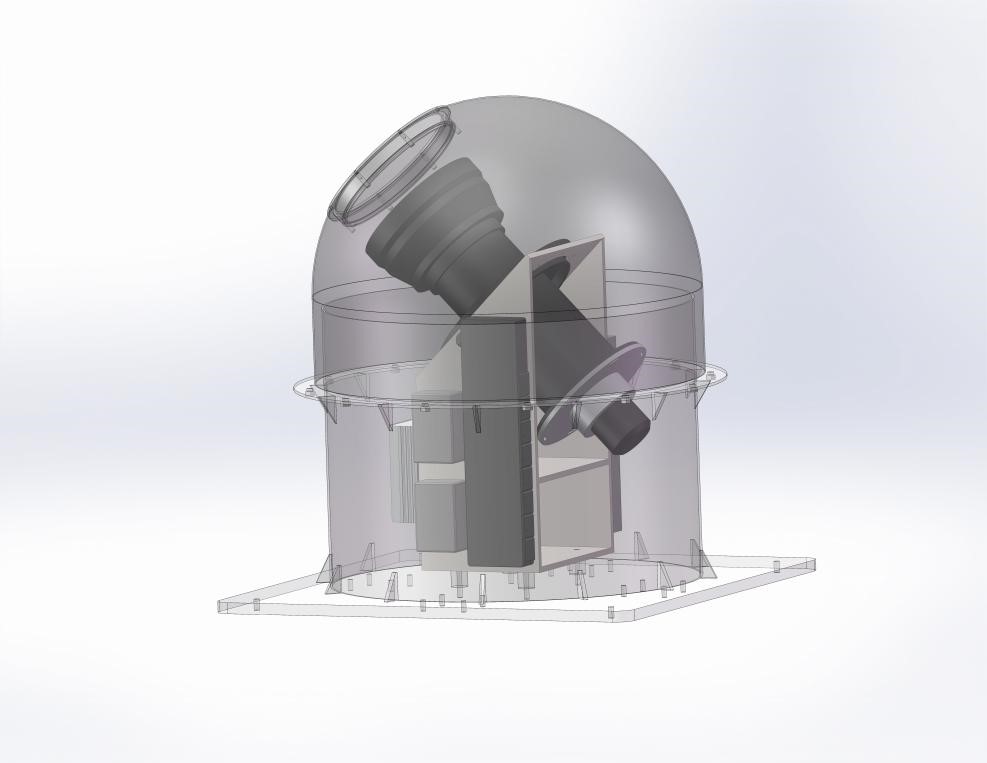}
    \caption{The structure of AT-Proto, showing key components, including the temperature-controlled dome, telescope system with a drift-scan CCD camera (180~mm aperture, 3k~×~3k pixel array), ITO-coated windows, and remote operation and control system.
}
    \label{fig:example1}
\end{figure}

The thermal control system integrates heat dissipation from detector electronics, computer components, and other heat sources, to maintain a stable internal temperature. The dome’s interior is regulated near a preset temperature (initially set at 5${\deg}$C), eliminating the need for cryogenic-resistant electronic components and frequent focus adjustments (see Figure ~\ref{fig:example1}).

For heating, the dome uses lightweight aerospace-grade polyimide film heating modules with high thermal efficiency. In full operation, these modules provide a maximum auxiliary heating power of 180~W, ensuring precise temperature control and stable temperature maintenance at approximately 5${\deg}$C.

\begin{figure*}[htbp]
    \centering
    \includegraphics[width=0.95\textwidth]{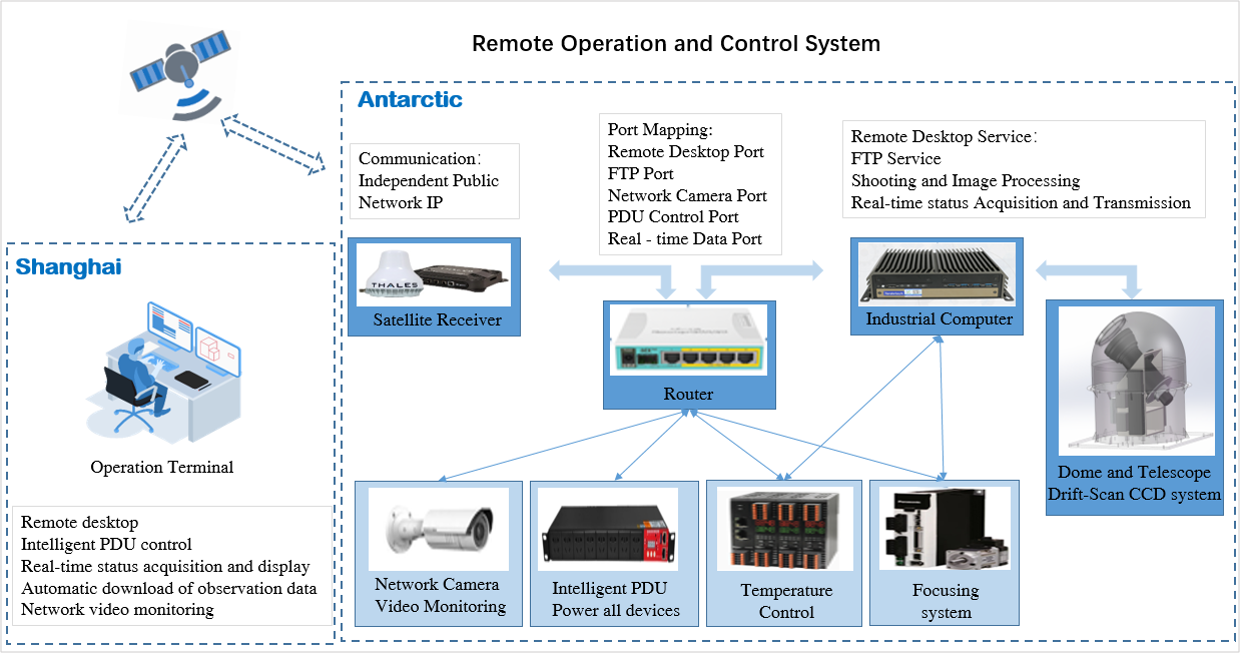}
   \caption{Structure diagram of the AT-Proto remote control system, showing two core components: The Antarctic field terminal (hardware controllers and satellite modems) and the Shanghai control center, which enables remote equipment status monitoring, start/stop control, on-site image processing, and transmission of preprocessed results from Shanghai. Remote data transmission is conducted via satellite terminals connecting to the global Internet.
    }
    \label{fig:example2}
\end{figure*}

\subsection{Drift-Scan CCD Camera}
For observations, a CCD camera with drift-scan capability is mounted at the focal plane of the optical telescope. Operating in drift-scan mode, the camera enables long-exposure integration of faint astronomical objects using charge integration via transfer \cite{2005PABei..23..304M}. This design eliminates the need for telescope tracking servos, significantly reducing development costs while enhancing system reliability. This makes it particularly suitable for harsh Antarctic environments and remote unattended operation.

The drift-scan CCD technique, also known as Time-Delay Integration (TDI), operates on the principle of progressive charge transfer. By synchronizing the charge transfer rate (controlled with timing circuitry) with the target’s angular drift velocity across the CCD, it achieves simultaneous charge accumulation and tracking \cite{2012AcASn..53..249S}. This produces well-defined stellar profiles for moving objects and enhances sensitivity to faint targets.

To accommodate variations in the angular drift velocities of celestial objects at different declinations, the CCD charge transfer speed is adjustable for the entire CCD. A customized control system was developed for the camera, supporting line transfer times from 10~ms to 1000~ms per line, meeting the prototype’s polar drift-scan requirements. The typical readout time is 12~s. Future iterations will extend this capability to higher-declination observations. Additionally, short-frame drift-scan was implemented to enable configurable exposure times.

For drift-scan observations, it is challenging to conduct observations at a unified drift-scan speed while maintaining the stability of star images across the entire FOV. This is particularly important for high-latitude regions, where the velocity difference between different latitudes is more significant. Given that our sky survey area lies in the high-latitude region (from -65${\deg}$ to -75${\deg}$), we have installed two detectors with independently controlled drift-scan speeds on the next-generation TianMu telescope. This design ensures that the star image offset of stars at different latitudes during a 30-second exposure is less than the Full Width at Half Maximum (FWHM). For the CCD positioned at the central declination of -72.5${\deg}$, the speed difference from the edge region at -75${\deg}$ is calculated as 15.04"/s × [cos(75${\deg}$) - cos(72.5${\deg}$)] = 0.62"/s. Considering the FWHM of ATSOP is 22.5", the maximum exposure time is determined to be 36.2~s. 

Given that the observation area of ATSOP is in a high-latitude region, we cannot use drift-scan technology for long-exposure observations to improve detection sensitivity. This is because of the curvature effect of star trails and the need to prevent severe image smearing at different latitudes. Consequently, we employ pixel-wise alignment and stacking techniques on CCD images obtained via multi-frame drift-scan. This process suppresses detector readout noise and cancels out random background interference, further improving the signal-to-noise ratio (SNR) of faint celestial objects. Once ATSOP is fully deployed, we will conduct pixel-wise alignment and stacking on CCD images acquired from different telescopes. This approach not only retains the advantage of large sky coverage offered by drift-scan but also breaks through the detection limit of a single device using stacking technology, providing higher-quality data support for research such as faint celestial object search and deep-space structure analysis.

\subsection{Remote Operation and Control System }
The Remote Operation and Control System is essential because both Taishan and Kunlun Stations in Antarctica operate uncrewed throughout winter, requiring all observation equipment to support remote control. This allows equipment status, start/stop operations, observation image processing, and transmission of pre-processed results to be monitored from Shanghai. As shown in Figure~\ref{fig:example2}, the system comprises the Antarctic field terminal (including hardware controllers and satellite modems) and the Shanghai control center. In Antarctica, satellite communication terminals (using Iridium protocols) connect to the global Internet for remote data transmission.

Given the constraints of satellite communications, the control system uses data content simplification, lightweight communication protocols, and lossless compression to reduce transmission volume, improving efficiency and enabling prompt detection of transient astronomical phenomena. This meets the requirements for unattended operation and low-bandwidth transmission.

\section{Observation Status of AT-Proto}

\subsection{Development Process of AT-Proto}
During the 2020~--~2022 construction phase of the ATSOP, the development of AT-Proto was successfully completed with support from the Center for Astronomical Mega-Science, Chinese Academy of Sciences, and independent projects of the Shanghai Astronomical Observatory. In August and November 2021, two low-temperature tests (at -60${\deg}$C) were conducted in Shanghai's cryogenic laboratory, to simulate Antarctic conditions, yielding crucial data. Analysis of these results led to further optimizations that enhanced the overall reliability of AT-Proto. From September to December 2021, optical system testing and installation were performed and completed.

Between January and March 2022, AT-Proto was transported to Inner Mongolia (where temperatures reach as low as -38${\deg}$C) for two months of astronomical observations. During these tests, the system operated stably, collecting substantial data that was successfully transmitted back to Shanghai. The thermal control system automatically activated when the cabin temperature fell below the setpoint of 5${\deg}$C, maintaining stable conditions. These field tests confirmed the reliability and readiness of AT-Proto for Antarctic deployment.

\begin{figure*}[htbp]
    \centering
    \includegraphics[width=0.85\textwidth]{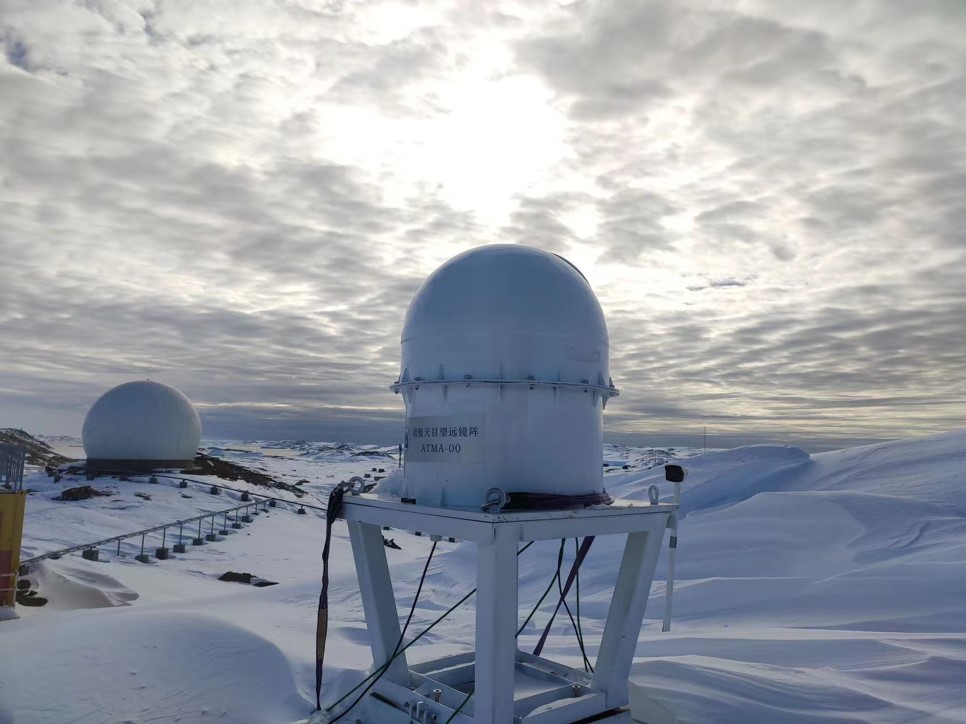 }
    \caption{AT-Proto at Zhongshan station.}
    \label{fig:example3}
\end{figure*}

AT-Proto was shipped to Zhongshan Station aboard the icebreaker Xuelong on October 31, 2022. Its integrated design facilitated smooth transportation and installation; after foundation work at the station, the equipment was hoisted into position and quickly commissioned through leveling and alignment adjustments (see Figure~\ref{fig:example3}). Aligned with Zhongshan Station's coordinates, the system's field of view points due north with a central declination of -25${\deg}$. During polar night, targets remain visible for approximately 40 minutes (the maximum duration of a target in one day) within the 10${\deg}$ field of view. Although the prototype’s continuous observation time for a specific target each day is relatively short, its high-frequency sampling still facilitates research on certain astronomical phenomena with short timescales (on the order of hours), such as low-mass X-ray binary systems \citep{2011ApJ...729....8Z}, $\delta$ Scuti variables \citep{2024AJ....168...94S}, and hot subdwarf binary systems \citep{2015A&A...576A..44K}.

\begin{figure}[htbp]
    \centering
    \includegraphics[width=0.5\textwidth]{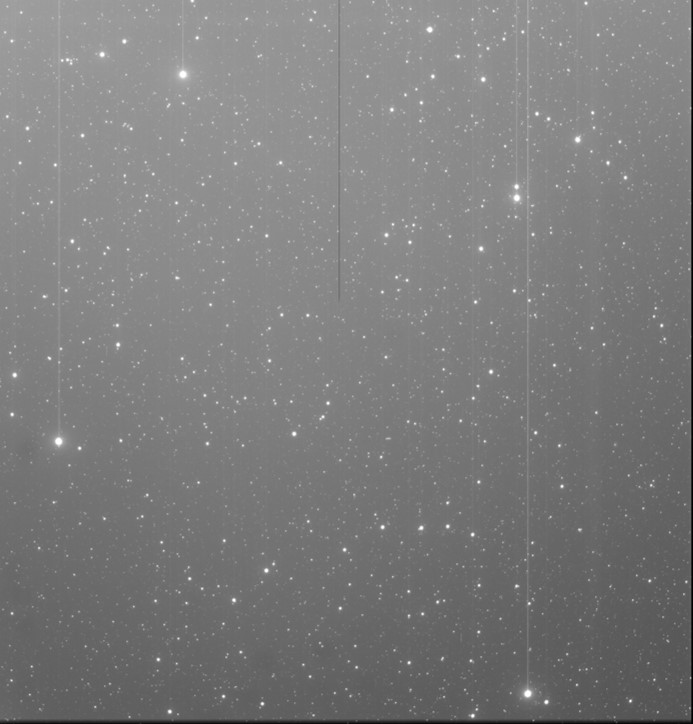 }
    \caption{First light image obtained by AT-Proto on February 19, 2023.}
    \label{fig:example4}
\end{figure}

First light was achieved on February 19, 2023, followed by trial observations from February 20 (see Figure ~\ref{fig:example4}). Without requiring focus adjustment, stellar images showed a 2-pixel FWHM, matching laboratory results. From March 2023, various parameters were tested under clear weather conditions, including exposure strategies, detection capabilities, photometric precision, and remote control functions. Successful continuous observations were conducted during polar night, from May 21 until July 15 (Niu et al. in prep). The 2023 observation season concluded on October 26, having collected 174,630 images (3.35~TB of data in total). In 2024, fault-free operations continued, acquiring 150,000 images between May 31 and September 5 (2.8~TB of data).

We define an image as "poor-quality" if it meets either of the following two criteria: The background sky brightness value exceeds 25,000~ADU (i.e. Analog-to-digital Units), or the number of stars identified in the field of view is less than 1,000. Such poor-quality images are
 caused by a bright lunar sky background and cloudy weather that reduces visibility. To investigate the number of days affected by cloudy weather, we screened the observation days with a median background sky brightness value of less than 25,000~ADU. The results showed that there were 6 such days in total, accounting for 11\% of the total polar night observation days. However, during the total 55-day polar night observation period, high-quality images were recorded on 46 days, accounting for 83\%.

\begin{figure*}[htbp]
    \centering
    \includegraphics[width=0.85\textwidth]{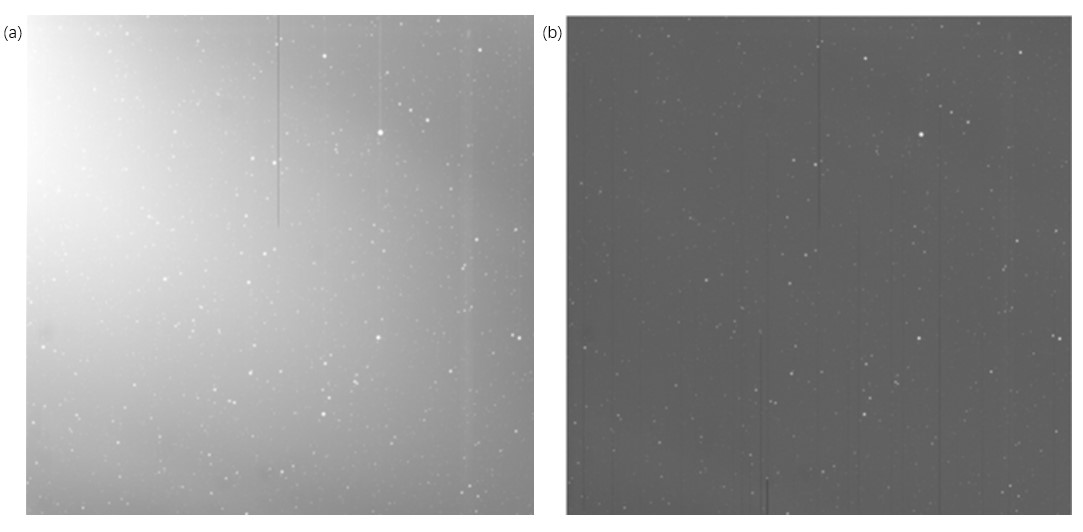}
    \caption{Sample CCD image of the prototype telescope taken from Zhongshan Station on May 1, 2023 (exposure time is 30 seconds). (a) Original image. (b) Processed image.}
    \label{fig:example5}
\end{figure*}

AT-Proto demonstrated remarkable stability during commissioning and testing, enduring extreme conditions of -37.3${\deg}$C minimum temperature and maximum wind speeds of 38.6~m/s, while maintaining excellent image quality. This confirms the advantages of drift-scan technology for polar astronomy applications.

\begin{figure*}[htbp]
    \centering
    \includegraphics[width=0.7\textwidth]{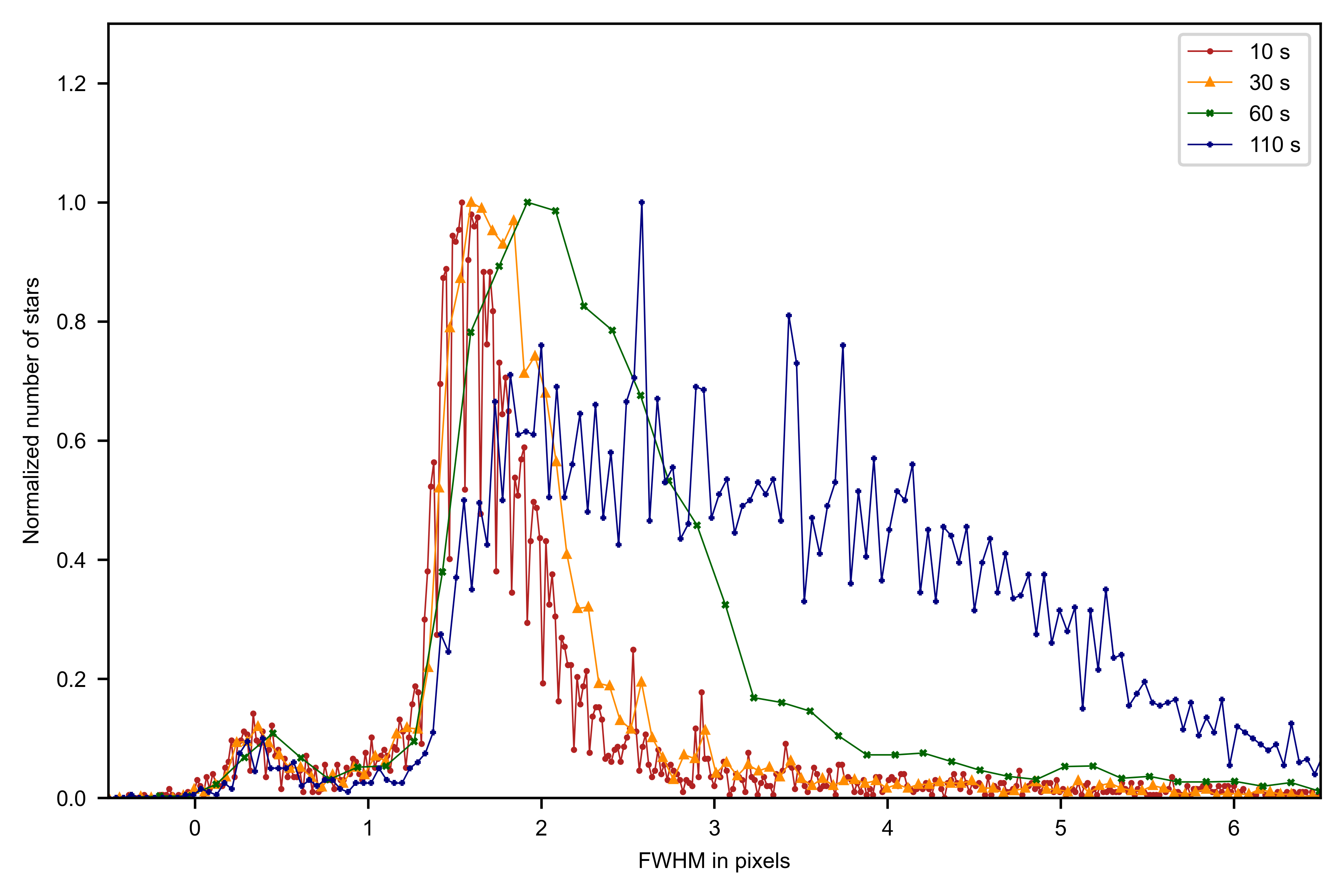 }
    \caption{FWHM distribution of stars captured at four different exposure times.}
    \label{fig:example6}
\end{figure*}

\begin{figure*}[htbp]
    \centering
    \includegraphics[width=0.85\textwidth]{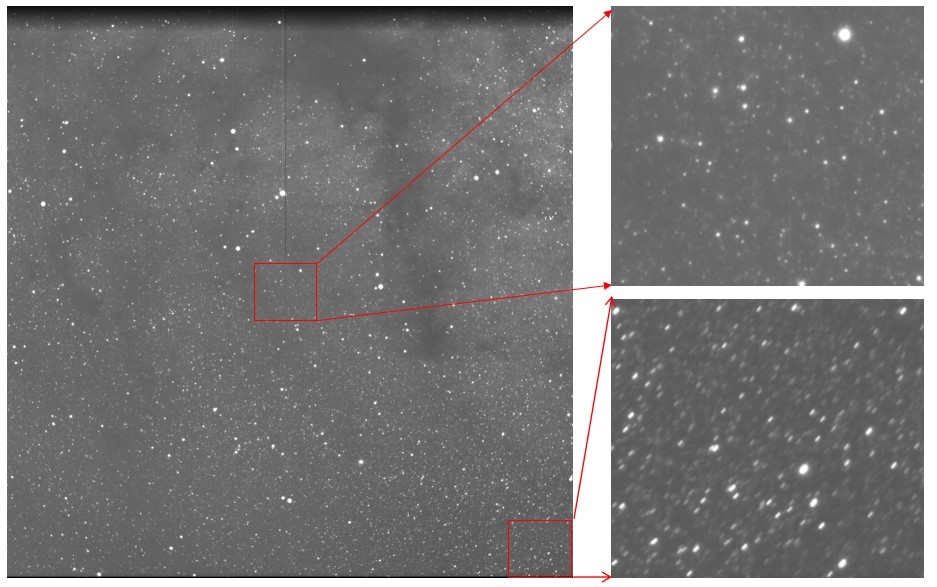  }
    \caption{The trailing effect with an exposure time of 110~s, demonstrating noticeable differences between the central and edge fields. The edge field clearly shows trailing effects, and FWHM measurements confirm significant degradation in image quality for 110~s exposures.}
    \label{fig:example7}
\end{figure*}

\subsection{Performance Testing of AT-Proto }
AT-Proto entered its trial observation phase on February 20, 2023. To evaluate the impact of different exposure times on image quality, observations were conducted using exposure times of 10~s, 30~s, 60~s, and 110~s. Selected images captured during nights with optimal observing conditions were transmitted via satellite to Shanghai for preliminary analysis.

Initial preprocessing was performed on the raw observational data (see the left panel of Figure~\ref{fig:example5}, including background subtraction, dark current removal, and smear correction. The background and dark current data were obtained during the prototype's laboratory testing at Shanghai Astronomical Observatory, with the CCD readout noise calculated as 14.4~$\pm$~2.4~e$^-$ from a series of background frames. The laboratory-measured dark current value was 0.27~$\pm$~0.06~e$^-$/pixel/s. Smear values were determined by measuring the flux in the upper and lower smear regions of each original observation image.

Background illumination was subsequently modeled and removed. The background sources included moonlight, ground reflections, lights from nearby polar stations, and stray light variations across CCD regions caused by long readout times (10~--~20~s). Following the Kepler satellite's image processing methodology \cite{2010ApJ...713L..79K}, we extracted fluxes from 5,600 uniformly spaced pixels along both horizontal and vertical directions for each image, then applied a two-dimensional fourth-order polynomial fit. The resulting model enabled the calculation and removal of background flux at any position on the CCD. The result is shown in the right panel of Figure~\ref{fig:example5}.

For images captured with four different exposure times (10~s, 30~s, 60~s, and 110~s) on the same night (April 30, 2023), we measured the FWHM of stars using SExtractor, with results shown in Figure~\ref{fig:example6}. The peak FWHM values for 10~s and 30~s exposures were 1.4 and 1.6 pixels (with a spatial scale of 11.25" per pixel), respectively, with most stars distributed between 1.2 and 2.5 pixels. These results perfectly matched laboratory optical system tests, confirming that observations could proceed without refocusing and validating the feasibility of the thermal control system and integrated structural design. For 60~s exposures, the peak FWHM was 2 pixels, with most stars between 1.2 and 3 pixels. For 110~s exposures, the maximum FWHM reached 6 pixels.

\begin{figure*}[htbp]
    \centering
    \includegraphics[width=0.95\textwidth]{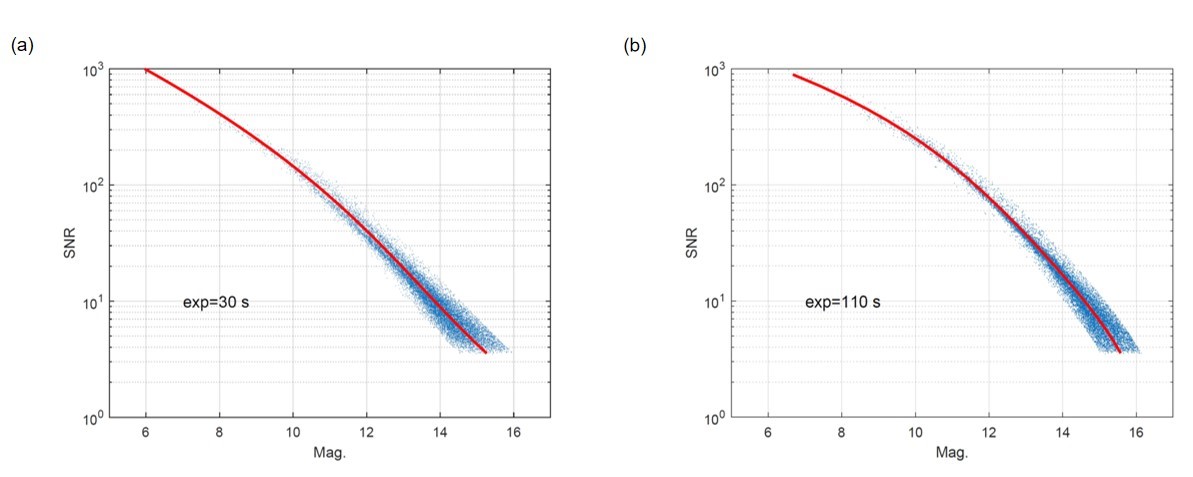}
    \caption{Detection capability of AT-Proto at Zhongshan Station, withno filter and magnitudes calibrated to the Gaia g-band.  (a) 30 second exposure time. (b) 110 second exposure time.}
    \label{fig:example8}
\end{figure*}
 
 \begin{figure*}[htbp]
    \centering
    \includegraphics[width=0.95\textwidth]{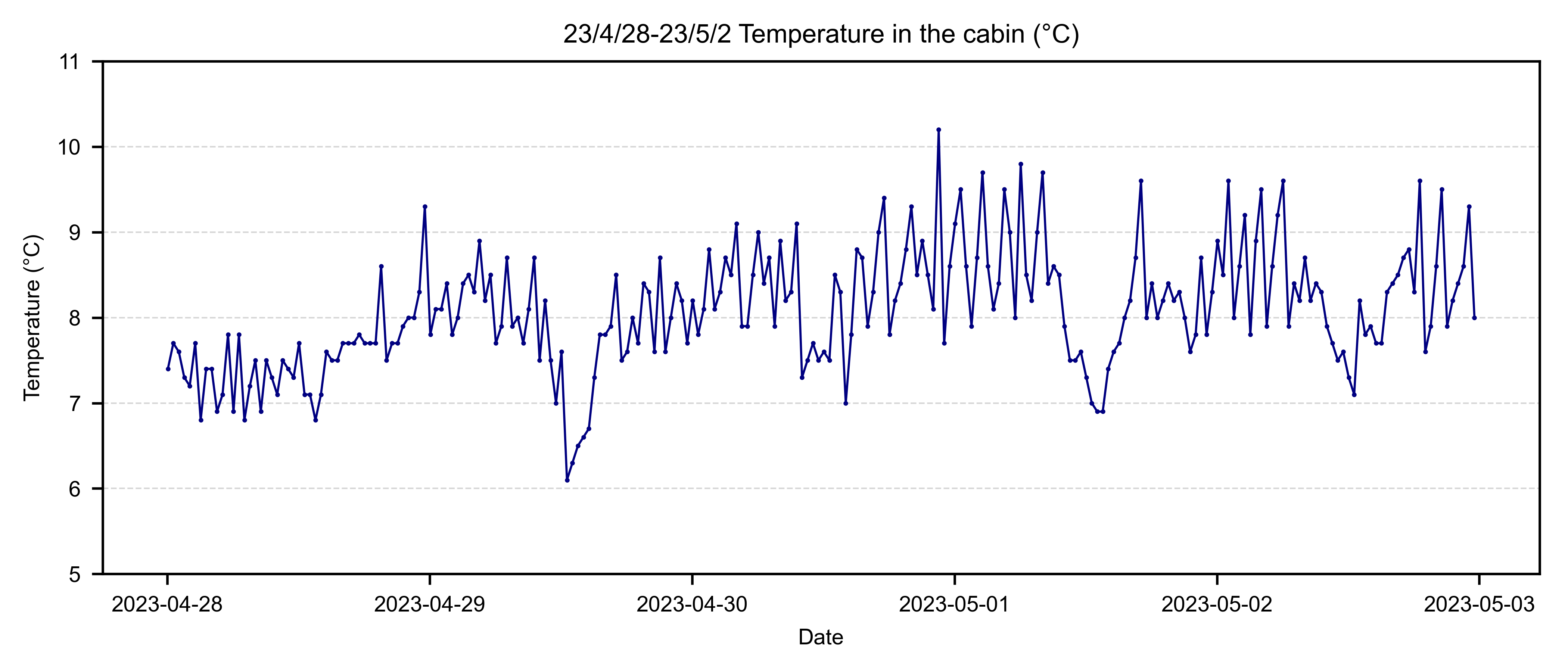 }
    \caption{Internal temperature inside the dome between April 28 and May 2, 2023.}
    \label{fig:example9}
\end{figure*}
  
The trailing effect at the field edges becomes increasingly severe with longer exposure times. This is partiall because, while the prototype's CCD drift-scan speed is set according to the linear velocity of stars at the field center of 13.63"/s at -25${\deg}$ declination, stars at different declinations have varying velocities -- 14.13"/s at -20${\deg}$ declination yields a difference of 0.5"/s. This results in accumulated positional discrepancies of 5" (0.5 pixels), 15" (1.4 pixels), 30" (2.8 pixels), and 55" (5 pixels) along right ascension for 10~s, 30~s, 60~s, and 110~s exposures respectively. The other main factor is that, because stars follow declination circles while drift-scan assumes linear motion, field rotation effects during prolonged exposures elongate stellar images at the field edges. As shown in Figure~\ref{fig:example7} for a 110~s exposure, the star images show noticeable differences between the central and edge fields, with a 5-pixel elongation caused by field rotation. The edge field clearly shows trailing effects, and FWHM measurements confirm significant degradation in image quality for 110~s exposures.

Given that typical star images span approximately 5 pixels, limiting elongation to this range ensures an optimal signal-to-noise ratio. We considered both factors, together with our analysis finding that 30~s exposures provide balanced photometric precision for both bright and faint sources while being less prone to saturation under bright auroras or full moonlight. Consequently, we selected 30~s as the optimal exposure time for sky surveys. We conducted a preliminary analysis of the detection capability of AT-Proto. The red line in Figure~\ref{fig:example8} shows the fitted curve between stellar magnitude and signal-to-noise ratio (SNR). Figure~\ref{fig:example8}(a) presents the magnitude distribution of stars detected in 30~s drift-scan CCD images during moonless nights, demonstrating that stars of 14.8 magnitudes (g-band) achieve SNR~$>$~5.0. Figure~\ref{fig:example8}(b) shows the results of 110~s exposures under similar conditions, with detectability limits reaching 15.3 magnitudes (g-band) at SNR~$>$~5.0. Stacking multiple images can further enhance both SNR and detection sensitivity, typically yielding an improvement of approximately 1.2 magnitudes when combining 10 frames. Future upgrades employing CCD cameras with higher quantum efficiency, lower dark current, and reduced readout noise will provide additional gains in detection performance.

To enhance the stability and safety of the prototype in polar conditions, based on extensive field tests at Zhongshan Station, we increased the temperature setpoint of the dome to 8${\deg}$C. This adjustment effectively prevents condensation on the inner window surface of the dome. As shown in Figure~\ref{fig:example9}, the dome temperature remains stable near the setpoint, with fluctuations below 2${\deg}$C. Since the stellar image quality fully meets observational requirements, no optical refocusing is needed. Moreover, this controlled cabin temperature effectively extends the lifespan of electronic components.

\section{Summary}

ATSOP uses the unique astronomical observation conditions of Antarctica, with core capabilities in high-cadence, uninterrupted, and large-sky-area staring sky surveys. By deploying an array of 30 small-aperture, large-field-of-view telescopes, it will carry out a rapid survey of minute-to-hour-scale optical transient sources and small celestial bodies with high orbital inclination in the southern celestial hemisphere.

The prototype telescope, AT-Proto, based on drift-scan CCD technology, has successfully operated in polar conditions for over two years, undergoing comprehensive testing of its observation modes. To minimize lunar background interference, the prototype was rotated by 180${\deg}$ in March 2025, switching its observation field from northern to southern skies. The equipment was re-hoisted and reinstalled with simple, efficient procedures. During Antarctic polar nights, the prototype acquired substantial observational data without any systemic failures, demonstrating excellent stability and reliability. These results confirm that the prototype's design effectively addresses the persistent technical issues and frequent malfunctions that have long been detrimental to Antarctic astronomical telescopes.

\section*{Acknowledgments}
This work is supported by the National Natural Science Foundation of China (NSFC) through grants 12090040 and 12090042, science research grants from the China Manned Space Project (grant NO. CMS-CSST-2025-A19), the Science and Technology Commission of Shanghai Municipality (Grant No.22dz1202400) and the Program of Shanghai Academic/Technology Research Leader.

\section*{Author Contributions}
Dan Zhou was responsible for designing the AT-Proto development plan, managing the instrument development, and drafting the original manuscript. Jing Zhong contributed to the conceptualization of the study, participated in drafting the original manuscript, and conducted the literature review. Hui Zhang provided research support and critically reviewed the manuscript. Jie Zhu, Lixin Zheng, Yu Yong, Jianjun Cao, Tang Zhenghong, Guoping Guo, Congcong Zhang, and Lurun Shen, as members of the AT-Proto research and development team, participated in all stages of the prototype development. Yaoxinyu contributed to data analysis. Peng Jiang, Xiang Pan, Chenwei Yang, and Tuo Ji actively assisted in the progress of the project and were responsible for the installation and operation of the AT-Proto at Zhongshan Station in Antarctica. Jianchun Shi, Xian Shi, Hengxiao Guo, Zhen Yan, Shiyin Shen, Donghai Zhao,  Liang Chen, Jianeng Zhou, Minfeng Gu, Fuguo Xie, Wenbiao Han, Jinliang Hou, Bixuan Zhao, Wenwen Zuo, Chun Xu, Zhengyi Shao, Lei Hao, and Jian Fu, as members of the ATSOP scientific working group, participated in the demonstration and refinement of scientific objectives. All authors read and approved the final manuscript.

\bibliography{ati}{}

@ARTICLE{2025RAA....25d4009H,
       author = {{Han}, Henggeng and {Huang}, Yang and {Wang}, Beichuan and {Sun}, Yongkang and {Wang}, Cunshi and {Li}, Zhirui and {Jin}, Junjie and {Sun}, Ningchen and {Xiao}, Kai and {He}, Min and {Gu}, Hongrui and {Niu}, Zexi and {Wu}, Hong and {Liu}, Jifeng},
        title = "{The Mini-SiTian Array: White Paper}",
      journal = {Research in Astronomy and Astrophysics},
         year = 2025,
        month = apr,
       volume = {25},
       number = {4},
          eid = {044009},
        pages = {044009},
          doi = {10.1088/1674-4527/adc791},
archivePrefix = {arXiv},
       eprint = {2504.01610},
 primaryClass = {astro-ph.IM},
       adsurl = {https://ui.adsabs.harvard.edu/abs/2025RAA....25d4009H}
}

@INPROCEEDINGS{2002SPIE.4836...10T,
       author = {{Tyson}, J. Anthony},
        title = "{Large Synoptic Survey Telescope: Overview}",
    booktitle = {Survey and Other Telescope Technologies and Discoveries},
         year = 2002,
       editor = {{Tyson}, J. Anthony and {Wolff}, Sidney},
       series = {Society of Photo-Optical Instrumentation Engineers (SPIE) Conference Series},
       volume = {4836},
        month = dec,
        pages = {10-20},
          doi = {10.1117/12.456772},
archivePrefix = {arXiv},
       eprint = {astro-ph/0302102},
 primaryClass = {astro-ph},
       adsurl = {https://ui.adsabs.harvard.edu/abs/2002SPIE.4836...10T}
}

@ARTICLE{2024FrASS..1104616H,
       author = {{Howell}, Steve B. and {Howell}, D. Andrew and {Street}, R.~A. and {Soares-Furtado}, Melinda and {Jackson}, Brian and {Greene}, Thomas P.},
        title = "{The dynamic universe: realizing the potential of classical time domain and multimessenger astrophysics}",
      journal = {Frontiers in Astronomy and Space Sciences},
         year = 2024,
        month = jan,
       volume = {11},
          eid = {1304616},
        pages = {1304616},
          doi = {10.3389/fspas.2024.1304616},
       adsurl = {https://ui.adsabs.harvard.edu/abs/2024FrASS..1104616H}
}

@ARTICLE{2010ApJ...716..781K,
       author = {{Katz}, Boaz and {Budnik}, Ran and {Waxman}, Eli},
        title = "{Fast Radiation Mediated Shocks and Supernova Shock Breakouts}",
      journal = {\apj},
         year = 2010,
        month = jun,
       volume = {716},
       number = {1},
        pages = {781-791},
          doi = {10.1088/0004-637X/716/1/781},
archivePrefix = {arXiv},
       eprint = {0902.4708},
 primaryClass = {astro-ph.HE},
       adsurl = {https://ui.adsabs.harvard.edu/abs/2010ApJ...716..781K}
}

@ARTICLE{2013ApJ...778..164A,
       author = {{Adams}, Scott M. and {Kochanek}, C.~S. and {Beacom}, John F. and {Vagins}, Mark R. and {Stanek}, K.~Z.},
        title = "{Observing the Next Galactic Supernova}",
      journal = {\apj},
         year = 2013,
        month = dec,
       volume = {778},
       number = {2},
          eid = {164},
        pages = {164},
          doi = {10.1088/0004-637X/778/2/164},
archivePrefix = {arXiv},
       eprint = {1306.0559},
 primaryClass = {astro-ph.HE},
       adsurl = {https://ui.adsabs.harvard.edu/abs/2013ApJ...778..164A}
}

@ARTICLE{2015JHEAp...7..148K,
       author = {{Komossa}, S.},
        title = "{Tidal disruption of stars by supermassive black holes: Status of observations}",
      journal = {Journal of High Energy Astrophysics},
         year = 2015,
        month = sep,
       volume = {7},
        pages = {148-157},
          doi = {10.1016/j.jheap.2015.04.006},
archivePrefix = {arXiv},
       eprint = {1505.01093},
 primaryClass = {astro-ph.HE},
       adsurl = {https://ui.adsabs.harvard.edu/abs/2015JHEAp...7..148K}
}

@BOOK{2008clno.book.....B,
       author = {{Bode}, Michael F. and {Evans}, Aneurin},
        title = "{Classical Novae}",
         year = 2008,
       volume = {43},
          doi = {10.1017/CBO9780511536168},
       adsurl = {https://ui.adsabs.harvard.edu/abs/2008clno.book.....B}
}

@INPROCEEDINGS{2008SPIE.7012E..4GY,
       author = {{Yuan}, Xiangyan and {Cui}, Xiangqun and {Liu}, Genrong and {Zhai}, Fengxiang and {Gong}, Xuefei and {Zhang}, Ru and {Xia}, Lirong and {Hu}, Jingyao and {Lawrence}, J.~S. and {Yan}, Jun and {Storey}, John W.~V. and {Wang}, Lifan and {Feng}, Longlong and {Ashley}, Michael C.~B. and {Zhou}, Xu and {Jiang}, Zhaoji and {Zhu}, Zhenxi},
        title = "{Chinese Small Telescope ARray (CSTAR) for Antarctic Dome A}",
    booktitle = {Ground-based and Airborne Telescopes II},
         year = 2008,
       editor = {{Stepp}, Larry M. and {Gilmozzi}, Roberto},
       series = {Society of Photo-Optical Instrumentation Engineers (SPIE) Conference Series},
       volume = {7012},
        month = jul,
          eid = {70124G},
        pages = {70124G},
          doi = {10.1117/12.788748},
       adsurl = {https://ui.adsabs.harvard.edu/abs/2008SPIE.7012E..4GY}
}

@INPROCEEDINGS{2014SPIE.9145E..0FY,
       author = {{Yuan}, Xiangyan and {Cui}, Xiangqun and {Gu}, Bozhong and {Yang}, Shihai and {Du}, Fujia and {Li}, Xiaoyan and {Wang}, Daxing and {Li}, Xinnan and {Gong}, Xuefei and {Wen}, Haikun and {Li}, Zhengyang and {Lu}, Haiping and {Xu}, Lingzhe and {Zhang}, Ru and {Zhang}, Yi and {Wang}, Lifan and {Shang}, Zhaohui and {Hu}, Yi and {Ma}, Bin and {Liu}, Qiang and {Wei}, Peng},
        title = "{The AST3 project: Antarctic Survey Telescopes for Dome A}",
    booktitle = {Ground-based and Airborne Telescopes V},
         year = 2014,
       editor = {{Stepp}, Larry M. and {Gilmozzi}, Roberto and {Hall}, Helen J.},
       series = {Society of Photo-Optical Instrumentation Engineers (SPIE) Conference Series},
       volume = {9145},
        month = jul,
          eid = {91450F},
        pages = {91450F},
          doi = {10.1117/12.2055624},
       adsurl = {https://ui.adsabs.harvard.edu/abs/2014SPIE.9145E..0FY}
}

@ARTICLE{2020MNRAS.497.1925G,
       author = {{Gomez}, Sebastian and {Nicholl}, Matt and {Short}, Philip and {Margutti}, Raffaella and {Alexander}, Kate D. and {Blanchard}, Peter K. and {Berger}, Edo and {Eftekhari}, Tarraneh and {Schulze}, Steve and {Anderson}, Joseph and {Arcavi}, Iair and {Chornock}, Ryan and {Cowperthwaite}, Philip S. and {Galbany}, Llu{\'\i}s and {Herzog}, Laura J. and {Hiramatsu}, Daichi and {Hosseinzadeh}, Griffin and {Laskar}, Tanmoy and {M{\"u}ller Bravo}, Tom{\'a}s E. and {Patton}, Locke and {Terreran}, Giacomo},
        title = "{The Tidal Disruption Event AT 2018hyz II: Light-curve modelling of a partially disrupted star}",
      journal = {\mnras},
         year = 2020,
        month = sep,
       volume = {497},
       number = {2},
        pages = {1925-1934},
          doi = {10.1093/mnras/staa2099},
archivePrefix = {arXiv},
       eprint = {2003.05469},
 primaryClass = {astro-ph.HE},
       adsurl = {https://ui.adsabs.harvard.edu/abs/2020MNRAS.497.1925G}
}

@ARTICLE{2013A&A...557A..12Z,
       author = {{Zaninoni}, E. and {Bernardini}, M.~G. and {Margutti}, R. and {Oates}, S. and {Chincarini}, G.},
        title = "{Gamma-ray burst optical light-curve zoo: comparison with X-ray observations}",
      journal = {\aap},
         year = 2013,
        month = sep,
       volume = {557},
          eid = {A12},
        pages = {A12},
          doi = {10.1051/0004-6361/201321221},
archivePrefix = {arXiv},
       eprint = {1303.6924},
 primaryClass = {astro-ph.HE},
       adsurl = {https://ui.adsabs.harvard.edu/abs/2013A&A...557A..12Z}
}

@ARTICLE{2023PASP..135c4102K,
       author = {{Kov{\'a}cs-Stermeczky}, Zs{\'o}fia V. and {Vink{\'o}}, J{\'o}zsef},
        title = "{Comparison of Different Tidal Disruption Event Light Curve Models with TiDE, a New Modular Open Source Code}",
      journal = {\pasp},
         year = 2023,
        month = mar,
       volume = {135},
       number = {1045},
          eid = {034102},
        pages = {034102},
          doi = {10.1088/1538-3873/acb9bb},
archivePrefix = {arXiv},
       eprint = {2302.08441},
 primaryClass = {astro-ph.HE},
       adsurl = {https://ui.adsabs.harvard.edu/abs/2023PASP..135c4102K}
}

@ARTICLE{2008Natur.455..183R,
       author = {{Racusin}, J.~L. and {Karpov}, S.~V. and {Sokolowski}, M. and {Granot}, J. and {Wu}, X.~F. and {Pal'Shin}, V. and {Covino}, S. and {van der Horst}, A.~J. and {Oates}, S.~R. and {Schady}, P. and {Smith}, R.~J. and {Cummings}, J. and {Starling}, R.~L.~C. and {Piotrowski}, L.~W. and {Zhang}, B. and {Evans}, P.~A. and {Holland}, S.~T. and {Malek}, K. and {Page}, M.~T. and {Vetere}, L. and {Margutti}, R. and {Guidorzi}, C. and {Kamble}, A.~P. and {Curran}, P.~A. and {Beardmore}, A. and {Kouveliotou}, C. and {Mankiewicz}, L. and {Melandri}, A. and {O'Brien}, P.~T. and {Page}, K.~L. and {Piran}, T. and {Tanvir}, N.~R. and {Wrochna}, G. and {Aptekar}, R.~L. and {Barthelmy}, S. and {Bartolini}, C. and {Beskin}, G.~M. and {Bondar}, S. and {Bremer}, M. and {Campana}, S. and {Castro-Tirado}, A. and {Cucchiara}, A. and {Cwiok}, M. and {D'Avanzo}, P. and {D'Elia}, V. and {Della Valle}, M. and {de Ugarte Postigo}, A. and {Dominik}, W. and {Falcone}, A. and {Fiore}, F. and {Fox}, D.~B. and {Frederiks}, D.~D. and {Fruchter}, A.~S. and {Fugazza}, D. and {Garrett}, M.~A. and {Gehrels}, N. and {Golenetskii}, S. and {Gomboc}, A. and {Gorosabel}, J. and {Greco}, G. and {Guarnieri}, A. and {Immler}, S. and {Jelinek}, M. and {Kasprowicz}, G. and {La Parola}, V. and {Levan}, A.~J. and {Mangano}, V. and {Mazets}, E.~P. and {Molinari}, E. and {Moretti}, A. and {Nawrocki}, K. and {Oleynik}, P.~P. and {Osborne}, J.~P. and {Pagani}, C. and {Pandey}, S.~B. and {Paragi}, Z. and {Perri}, M. and {Piccioni}, A. and {Ramirez-Ruiz}, E. and {Roming}, P.~W.~A. and {Steele}, I.~A. and {Strom}, R.~G. and {Testa}, V. and {Tosti}, G. and {Ulanov}, M.~V. and {Wiersema}, K. and {Wijers}, R.~A.~M.~J. and {Winters}, J.~M. and {Zarnecki}, A.~F. and {Zerbi}, F. and {M{\'e}sz{\'a}ros}, P. and {Chincarini}, G. and {Burrows}, D.~N.},
        title = "{Broadband observations of the naked-eye {\ensuremath{\gamma}}-ray burst GRB080319B}",
      journal = {\nat},
         year = 2008,
        month = sep,
       volume = {455},
       number = {7210},
        pages = {183-188},
          doi = {10.1038/nature07270},
archivePrefix = {arXiv},
       eprint = {0805.1557},
 primaryClass = {astro-ph},
       adsurl = {https://ui.adsabs.harvard.edu/abs/2008Natur.455..183R}
}

@ARTICLE{2020MNRAS.494.2191N,
       author = {{Namouni}, F. and {Morais}, M.~H.~M.},
        title = "{An interstellar origin for high-inclination Centaurs}",
      journal = {\mnras},
         year = 2020,
        month = may,
       volume = {494},
       number = {2},
        pages = {2191-2199},
          doi = {10.1093/mnras/staa712},
archivePrefix = {arXiv},
       eprint = {2004.10510},
 primaryClass = {astro-ph.EP},
       adsurl = {https://ui.adsabs.harvard.edu/abs/2020MNRAS.494.2191N}
}

@ARTICLE{2020Natur.583..771M,
       author = {{Ma}, Bin and {Shang}, Zhaohui and {Hu}, Yi and {Hu}, Keliang and {Wang}, Yongjiang and {Yang}, Xu and {Ashley}, Michael C.~B. and {Hickson}, Paul and {Jiang}, Peng},
        title = "{Night-time measurements of astronomical seeing at Dome A in Antarctica}",
      journal = {\nat},
         year = 2020,
        month = jul,
       volume = {583},
       number = {7818},
        pages = {771-774},
          doi = {10.1038/s41586-020-2489-0},
archivePrefix = {arXiv},
       eprint = {2007.15365},
 primaryClass = {astro-ph.IM},
       adsurl = {https://ui.adsabs.harvard.edu/abs/2020Natur.583..771M}
}

@ARTICLE{2012A&ARv..20...56C,
       author = {{Caselli}, Paola and {Ceccarelli}, Cecilia},
        title = "{Our astrochemical heritage}",
      journal = {\aapr},
         year = 2012,
        month = oct,
       volume = {20},
          eid = {56},
        pages = {56},
          doi = {10.1007/s00159-012-0056-x},
archivePrefix = {arXiv},
       eprint = {1210.6368},
 primaryClass = {astro-ph.GA},
       adsurl = {https://ui.adsabs.harvard.edu/abs/2012A&ARv..20...56C}
}

@ARTICLE{2021AnABC..93..628L,
       author = {{Liu}, Jifeng and {Soria}, Roberto and {Wu}, Xue-Feng and {Wu}, Hong and {Shang}, Zhaohui},
        title = "{The SiTian Project}",
      journal = {Anais da Academia Brasileira de Ciencias},
         year = 2021,
        month = apr,
       volume = {93},
          eid = {20200628},
        pages = {20200628},
          doi = {10.1590/0001-3765202120200628},
archivePrefix = {arXiv},
       eprint = {2006.01844},
 primaryClass = {astro-ph.IM},
       adsurl = {https://ui.adsabs.harvard.edu/abs/2021AnABC..93..628L}
}

@ARTICLE{2011ApJ...729....8Z,
       author = {{Zhong}, Jing and {Wang}, Zhongxiang},
        title = "{The Likely Orbital Period of the Ultracompact Low-mass X-ray Binary 2S 0918-549}",
      journal = {\apj},
         year = 2011,
        month = mar,
       volume = {729},
       number = {1},
          eid = {8},
        pages = {8},
          doi = {10.1088/0004-637X/729/1/8},
archivePrefix = {arXiv},
       eprint = {1006.3980},
 primaryClass = {astro-ph.HE},
       adsurl = {https://ui.adsabs.harvard.edu/abs/2011ApJ...729....8Z}
}

@ARTICLE{2002MNRAS.329..897H,
       author = {{Hurley}, Jarrod R. and {Tout}, Christopher A. and {Pols}, Onno R.},
        title = "{Evolution of binary stars and the effect of tides on binary populations}",
      journal = {\mnras},
         year = 2002,
        month = feb,
       volume = {329},
       number = {4},
        pages = {897-928},
          doi = {10.1046/j.1365-8711.2002.05038.x},
archivePrefix = {arXiv},
       eprint = {astro-ph/0201220},
 primaryClass = {astro-ph},
       adsurl = {https://ui.adsabs.harvard.edu/abs/2002MNRAS.329..897H}
}

@ARTICLE{2010ApJS..190....1R,
       author = {{Raghavan}, Deepak and {McAlister}, Harold A. and {Henry}, Todd J. and {Latham}, David W. and {Marcy}, Geoffrey W. and {Mason}, Brian D. and {Gies}, Douglas R. and {White}, Russel J. and {ten Brummelaar}, Theo A.},
        title = "{A Survey of Stellar Families: Multiplicity of Solar-type Stars}",
      journal = {\apjs},
         year = 2010,
        month = sep,
       volume = {190},
       number = {1},
        pages = {1-42},
          doi = {10.1088/0067-0049/190/1/1},
archivePrefix = {arXiv},
       eprint = {1007.0414},
 primaryClass = {astro-ph.SR},
       adsurl = {https://ui.adsabs.harvard.edu/abs/2010ApJS..190....1R}
}

@INPROCEEDINGS{2013EAS....64..269B,
       author = {{Bloemen}, S. and {Degroote}, P. and {Conroy}, K. and {Hambleton}, K.~M. and {Giammarco}, J.~M. and {Pablo}, H. and {Pr{\v{s}}a}, A.},
        title = "{Physics of Eclipsing Binaries: Modelling in the new era of ultra-high precision photometry}",
    booktitle = {EAS Publications Series},
         year = 2013,
       editor = {{Pavlovski}, K. and {Tkachenko}, A. and {Torres}, G.},
       series = {EAS Publications Series},
       volume = {64},
        month = feb,
        pages = {269-276},
          doi = {10.1051/eas/1364037},
       adsurl = {https://ui.adsabs.harvard.edu/abs/2013EAS....64..269B}
}

@ARTICLE{2002A&A...391..213K,
       author = {{Kim}, S. -L. and {Lee}, J.~W. and {Youn}, J. -H. and {Kwon}, S. -G. and {Kim}, C.},
        title = "{Photometric study of a pulsating component in the eclipsing binary Y Cam}",
      journal = {\aap},
         year = 2002,
        month = aug,
       volume = {391},
        pages = {213-218},
          doi = {10.1051/0004-6361:20020777},
       adsurl = {https://ui.adsabs.harvard.edu/abs/2002A&A...391..213K}
}

@ARTICLE{2025arXiv250620540O,
       author = {{Ouyed}, Rachid},
        title = "{Optical Flares in the Luminous Fast Blue Optical Transient AT2022tsd (``Tasmanian Devil'')}",
      journal = {arXiv e-prints},
         year = 2025,
        month = jun,
          eid = {arXiv:2506.20540},
        pages = {arXiv:2506.20540},
archivePrefix = {arXiv},
       eprint = {2506.20540},
 primaryClass = {astro-ph.HE},
       adsurl = {https://ui.adsabs.harvard.edu/abs/2025arXiv250620540O}
}

@ARTICLE{2010ApJ...713L.169C,
       author = {{Chaplin}, W.~J. and {Appourchaux}, T. and {Elsworth}, Y. and {Garc{\'\i}a}, R.~A. and {Houdek}, G. and {Karoff}, C. and {Metcalfe}, T.~S. and {Molenda-{\.Z}akowicz}, J. and {Monteiro}, M.~J.~P.~F.~G. and {Thompson}, M.~J. and {Brown}, T.~M. and {Christensen-Dalsgaard}, J. and {Gilliland}, R.~L. and {Kjeldsen}, H. and {Borucki}, W.~J. and {Koch}, D. and {Jenkins}, J.~M. and {Ballot}, J. and {Basu}, S. and {Bazot}, M. and {Bedding}, T.~R. and {Benomar}, O. and {Bonanno}, A. and {Brand{\~a}o}, I.~M. and {Bruntt}, H. and {Campante}, T.~L. and {Creevey}, O.~L. and {Di Mauro}, M.~P. and {Do{\v{g}}an}, G. and {Dreizler}, S. and {Eggenberger}, P. and {Esch}, L. and {Fletcher}, S.~T. and {Frandsen}, S. and {Gai}, N. and {Gaulme}, P. and {Handberg}, R. and {Hekker}, S. and {Howe}, R. and {Huber}, D. and {Korzennik}, S.~G. and {Lebrun}, J.~C. and {Leccia}, S. and {Martic}, M. and {Mathur}, S. and {Mosser}, B. and {New}, R. and {Quirion}, P. -O. and {R{\'e}gulo}, C. and {Roxburgh}, I.~W. and {Salabert}, D. and {Schou}, J. and {Sousa}, S.~G. and {Stello}, D. and {Verner}, G.~A. and {Arentoft}, T. and {Barban}, C. and {Belkacem}, K. and {Benatti}, S. and {Biazzo}, K. and {Boumier}, P. and {Bradley}, P.~A. and {Broomhall}, A. -M. and {Buzasi}, D.~L. and {Claudi}, R.~U. and {Cunha}, M.~S. and {D'Antona}, F. and {Deheuvels}, S. and {Derekas}, A. and {Garc{\'\i}a Hern{\'a}ndez}, A. and {Giampapa}, M.~S. and {Goupil}, M.~J. and {Gruberbauer}, M. and {Guzik}, J.~A. and {Hale}, S.~J. and {Ireland}, M.~J. and {Kiss}, L.~L. and {Kitiashvili}, I.~N. and {Kolenberg}, K. and {Korhonen}, H. and {Kosovichev}, A.~G. and {Kupka}, F. and {Lebreton}, Y. and {Leroy}, B. and {Ludwig}, H. -G. and {Mathis}, S. and {Michel}, E. and {Miglio}, A. and {Montalb{\'a}n}, J. and {Moya}, A. and {Noels}, A. and {Noyes}, R.~W. and {Pall{\'e}}, P.~L. and {Piau}, L. and {Preston}, H.~L. and {Roca Cort{\'e}s}, T. and {Roth}, M. and {Sato}, K.~H. and {Schmitt}, J. and {Serenelli}, A.~M. and {Silva Aguirre}, V. and {Stevens}, I.~R. and {Su{\'a}rez}, J.~C. and {Suran}, M.~D. and {Trampedach}, R. and {Turck-Chi{\`e}ze}, S. and {Uytterhoeven}, K. and {Ventura}, R. and {Wilson}, P.~A.},
        title = "{The Asteroseismic Potential of Kepler: First Results for Solar-Type Stars}",
      journal = {\apjl},
         year = 2010,
        month = apr,
       volume = {713},
       number = {2},
        pages = {L169-L175},
          doi = {10.1088/2041-8205/713/2/L169},
archivePrefix = {arXiv},
       eprint = {1001.0506},
 primaryClass = {astro-ph.SR},
       adsurl = {https://ui.adsabs.harvard.edu/abs/2010ApJ...713L.169C}
}

@ARTICLE{2018MNRAS.479.2777R,
       author = {{Rivera Sandoval}, L.~E. and {Wijnands}, R. and {Degenaar}, N. and {Cavecchi}, Y. and {Heinke}, C.~O. and {Cackett}, E.~M. and {Homan}, J. and {Altamirano}, D. and {Bahramian}, A. and {Sivakoff}, G.~R. and {Miller}, J.~M. and {Parikh}, A.~S.},
        title = "{Extreme quiescent variability of the transient neutron star low-mass X-ray binary EXO 1745-248 in Terzan 5}",
      journal = {\mnras},
         year = 2018,
        month = sep,
       volume = {479},
       number = {2},
        pages = {2777-2788},
          doi = {10.1093/mnras/sty1535},
archivePrefix = {arXiv},
       eprint = {1711.05876},
 primaryClass = {astro-ph.HE},
       adsurl = {https://ui.adsabs.harvard.edu/abs/2018MNRAS.479.2777R}
}

@ARTICLE{2014ApJ...797..121H,
       author = {{Hawley}, Suzanne L. and {Davenport}, James R.~A. and {Kowalski}, Adam F. and {Wisniewski}, John P. and {Hebb}, Leslie and {Deitrick}, Russell and {Hilton}, Eric J.},
        title = "{Kepler Flares. I. Active and Inactive M Dwarfs}",
      journal = {\apj},
         year = 2014,
        month = dec,
       volume = {797},
       number = {2},
          eid = {121},
        pages = {121},
          doi = {10.1088/0004-637X/797/2/121},
archivePrefix = {arXiv},
       eprint = {1410.7779},
 primaryClass = {astro-ph.SR},
       adsurl = {https://ui.adsabs.harvard.edu/abs/2014ApJ...797..121H}
}

@ARTICLE{2015AJ....149...84Z,
       author = {{Zong}, Weikai and {Fu}, Jian-Ning and {Niu}, Jia-Shu and {Charpinet}, S. and {Vauclair}, G. and {Ashley}, Michael C.~B. and {Cui}, Xiangqun and {Feng}, Longlong and {Gong}, Xuefei and {Lawrence}, Jon S. and {Luong-Van}, Daniel and {Liu}, Qiang and {Pennypacker}, Carl R. and {Wang}, Lingzhi and {Wang}, Lifan and {Yuan}, Xiangyan and {York}, Donald G. and {Zhou}, Xu and {Zhu}, Zhenxi and {Zhu}, Zonghong},
        title = "{Discovery of Multiple Pulsations in the New {\ensuremath{\delta}} Scuti Star HD 92277: Asteroseismology from Dome A, Antarctica}",
      journal = {\aj},
         year = 2015,
        month = feb,
       volume = {149},
       number = {2},
          eid = {84},
        pages = {84},
          doi = {10.1088/0004-6256/149/2/84},
       adsurl = {https://ui.adsabs.harvard.edu/abs/2015AJ....149...84Z}
}

@INPROCEEDINGS{2017EPJWC.16004003Z,
       author = {{Zong}, Weikai and {Charpinet}, St{\'e}phane and {Vauclair}, G{\'e}rard and {Giammichele}, Noemi and {Van Grootel}, Val{\'e}rie},
        title = "{Nonlinear asteroseismology: insight from amplitude and frequency modulations of oscillation modes in compact pulsators from Kepler photometry}",
    booktitle = {European Physical Journal Web of Conferences},
         year = 2017,
       series = {European Physical Journal Web of Conferences},
       volume = {160},
        month = oct,
          eid = {04003},
        pages = {04003},
          doi = {10.1051/epjconf/201716004003},
       adsurl = {https://ui.adsabs.harvard.edu/abs/2017EPJWC.16004003Z}
}

@ARTICLE{1958HDP....51..353L,
       author = {{Ledoux}, Paul and {Walraven}, Th{\'e}odore},
        title = "{Variable Stars.}",
      journal = {Handbuch der Physik},
         year = 1958,
        month = jan,
       volume = {51},
        pages = {353-604},
          doi = {10.1007/978-3-642-45908-5_6},
       adsurl = {https://ui.adsabs.harvard.edu/abs/1958HDP....51..353L}
}

@ARTICLE{2020ApJS..249...18C,
       author = {{Chen}, Xiaodian and {Wang}, Shu and {Deng}, Licai and {de Grijs}, Richard and {Yang}, Ming and {Tian}, Hao},
        title = "{The Zwicky Transient Facility Catalog of Periodic Variable Stars}",
      journal = {\apjs},
         year = 2020,
        month = jul,
       volume = {249},
       number = {1},
          eid = {18},
        pages = {18},
          doi = {10.3847/1538-4365/ab9cae},
archivePrefix = {arXiv},
       eprint = {2005.08662},
 primaryClass = {astro-ph.SR},
       adsurl = {https://ui.adsabs.harvard.edu/abs/2020ApJS..249...18C}
}

@BOOK{2003cvs..book.....W,
       author = {{Warner}, Brian},
        title = "{Cataclysmic Variable Stars}",
         year = 2003,
          doi = {10.1017/CBO9780511586491},
       adsurl = {https://ui.adsabs.harvard.edu/abs/2003cvs..book.....W}
}

@ARTICLE{2025RAA....25d4001H,
       author = {{Huang}, Yang and {Liu}, Jifeng and {Wu}, Hong and {Shang}, Zhaohui and {Luo}, Ali and {Hu}, Shaoming and {Cui}, Wenyuan and {Mao}, Yongna},
        title = "{The Mini-SiTian Array: A Pathfinder for the SiTian Project}",
      journal = {Research in Astronomy and Astrophysics},
         year = 2025,
        month = apr,
       volume = {25},
       number = {4},
          eid = {044001},
        pages = {044001},
          doi = {10.1088/1674-4527/adc795},
archivePrefix = {arXiv},
       eprint = {2504.01615},
 primaryClass = {astro-ph.IM},
       adsurl = {https://ui.adsabs.harvard.edu/abs/2025RAA....25d4001H}
}

@INPROCEEDINGS{2018SPIE10700E..1LL,
       author = {{Li}, Zhengyang and {Yuan}, Xiangyan and {Cui}, Xiangqun and {Wang}, Lifan and {Shang}, Zhaohui and {Du}, Fujia and {Gong}, Xuefei and {Gu}, Bozhong and {Hu}, Yi and {Jiang}, Peng and {Li}, Xiaoyan and {Lu}, Haiping and {Ma}, Bin and {Wei}, Fuhai and {Wen}, Haikun and {Xu}, Jin and {Yang}, Shihai and {Zhou}, Honyan},
        title = "{Introduction of Chinese Antarctic optical telescopes}",
    booktitle = {Ground-based and Airborne Telescopes VII},
         year = 2018,
       editor = {{Marshall}, Heather K. and {Spyromilio}, Jason},
       series = {Society of Photo-Optical Instrumentation Engineers (SPIE) Conference Series},
       volume = {10700},
        month = jul,
          eid = {107001L},
        pages = {107001L},
          doi = {10.1117/12.2309618},
       adsurl = {https://ui.adsabs.harvard.edu/abs/2018SPIE10700E..1LL}
}

@ARTICLE{2019ApJS..241...29Y,
       author = {{Yang}, Huiqin and {Liu}, Jifeng},
        title = "{The Flare Catalog and the Flare Activity in the Kepler Mission}",
      journal = {\apjs},
         year = 2019,
        month = apr,
       volume = {241},
       number = {2},
          eid = {29},
        pages = {29},
          doi = {10.3847/1538-4365/ab0d28},
archivePrefix = {arXiv},
       eprint = {1903.01056},
 primaryClass = {astro-ph.SR},
       adsurl = {https://ui.adsabs.harvard.edu/abs/2019ApJS..241...29Y}
}

@ARTICLE{2013ApJS..209....5S,
       author = {{Shibayama}, Takuya and {Maehara}, Hiroyuki and {Notsu}, Shota and {Notsu}, Yuta and {Nagao}, Takashi and {Honda}, Satoshi and {Ishii}, Takako T. and {Nogami}, Daisaku and {Shibata}, Kazunari},
        title = "{Superflares on Solar-type Stars Observed with Kepler. I. Statistical Properties of Superflares}",
      journal = {\apjs},
         year = 2013,
        month = nov,
       volume = {209},
       number = {1},
          eid = {5},
        pages = {5},
          doi = {10.1088/0067-0049/209/1/5},
archivePrefix = {arXiv},
       eprint = {1308.1480},
 primaryClass = {astro-ph.SR},
       adsurl = {https://ui.adsabs.harvard.edu/abs/2013ApJS..209....5S}
}

@ARTICLE{2008Natur.453..469S,
       author = {{Soderberg}, A.~M. and {Berger}, E. and {Page}, K.~L. and {Schady}, P. and {Parrent}, J. and {Pooley}, D. and {Wang}, X. -Y. and {Ofek}, E.~O. and {Cucchiara}, A. and {Rau}, A. and {Waxman}, E. and {Simon}, J.~D. and {Bock}, D.~C. -J. and {Milne}, P.~A. and {Page}, M.~J. and {Barentine}, J.~C. and {Barthelmy}, S.~D. and {Beardmore}, A.~P. and {Bietenholz}, M.~F. and {Brown}, P. and {Burrows}, A. and {Burrows}, D.~N. and {Byrngelson}, G. and {Cenko}, S.~B. and {Chandra}, P. and {Cummings}, J.~R. and {Fox}, D.~B. and {Gal-Yam}, A. and {Gehrels}, N. and {Immler}, S. and {Kasliwal}, M. and {Kong}, A.~K.~H. and {Krimm}, H.~A. and {Kulkarni}, S.~R. and {Maccarone}, T.~J. and {M{\'e}sz{\'a}ros}, P. and {Nakar}, E. and {O'Brien}, P.~T. and {Overzier}, R.~A. and {de Pasquale}, M. and {Racusin}, J. and {Rea}, N. and {York}, D.~G.},
        title = "{An extremely luminous X-ray outburst at the birth of a supernova}",
      journal = {\nat},
         year = 2008,
        month = may,
       volume = {453},
       number = {7194},
        pages = {469-474},
          doi = {10.1038/nature06997},
archivePrefix = {arXiv},
       eprint = {0802.1712},
 primaryClass = {astro-ph},
       adsurl = {https://ui.adsabs.harvard.edu/abs/2008Natur.453..469S}
}

@ARTICLE{2004RvMP...76.1143P,
       author = {{Piran}, Tsvi},
        title = "{The physics of gamma-ray bursts}",
      journal = {Reviews of Modern Physics},
         year = 2004,
        month = oct,
       volume = {76},
       number = {4},
        pages = {1143-1210},
          doi = {10.1103/RevModPhys.76.1143},
archivePrefix = {arXiv},
       eprint = {astro-ph/0405503},
 primaryClass = {astro-ph},
       adsurl = {https://ui.adsabs.harvard.edu/abs/2004RvMP...76.1143P}
}

@ARTICLE{2012ApJ...747...88N,
       author = {{Nakar}, Ehud and {Sari}, Re'em},
        title = "{Relativistic Shock Breakouts{\textemdash}A Variety of Gamma-Ray Flares: From Low-luminosity Gamma-Ray Bursts to Type Ia Supernovae}",
      journal = {\apj},
         year = 2012,
        month = mar,
       volume = {747},
       number = {2},
          eid = {88},
        pages = {88},
          doi = {10.1088/0004-637X/747/2/88},
archivePrefix = {arXiv},
       eprint = {1106.2556},
 primaryClass = {astro-ph.HE},
       adsurl = {https://ui.adsabs.harvard.edu/abs/2012ApJ...747...88N}
}

@ARTICLE{2016PhRvL.116f1102A,
       author = {{Abbott}, B.~P. and {Abbott}, R. and {Abbott}, T.~D. and {Abernathy}, M.~R. and {Acernese}, F. and {Ackley}, K. and {Adams}, C. and {Adams}, T. and {Addesso}, P. and {Adhikari}, R.~X. and {Adya}, V.~B. and {Affeldt}, C. and {Agathos}, M. and {Agatsuma}, K. and {Aggarwal}, N. and {Aguiar}, O.~D. and {Aiello}, L. and {Ain}, A. and {Ajith}, P. and {Allen}, B. and {Allocca}, A. and {Altin}, P.~A. and {Anderson}, S.~B. and {Anderson}, W.~G. and {Arai}, K. and {Arain}, M.~A. and {Araya}, M.~C. and {Arceneaux}, C.~C. and {Areeda}, J.~S. and {Arnaud}, N. and {Arun}, K.~G. and {Ascenzi}, S. and {Ashton}, G. and {Ast}, M. and {Aston}, S.~M. and {Astone}, P. and {Aufmuth}, P. and {Aulbert}, C. and {Babak}, S. and {Bacon}, P. and {Bader}, M.~K.~M. and {Baker}, P.~T. and {Baldaccini}, F. and {Ballardin}, G. and {Ballmer}, S.~W. and {Barayoga}, J.~C. and {Barclay}, S.~E. and {Barish}, B.~C. and {Barker}, D. and {Barone}, F. and {Barr}, B. and {Barsotti}, L. and {Barsuglia}, M. and {Barta}, D. and {Bartlett}, J. and {Barton}, M.~A. and {Bartos}, I. and {Bassiri}, R. and {Basti}, A. and {Batch}, J.~C. and {Baune}, C. and {Bavigadda}, V. and {Bazzan}, M. and {Behnke}, B. and {Bejger}, M. and {Belczynski}, C. and {Bell}, A.~S. and {Bell}, C.~J. and {Berger}, B.~K. and {Bergman}, J. and {Bergmann}, G. and {Berry}, C.~P.~L. and {Bersanetti}, D. and {Bertolini}, A. and {Betzwieser}, J. and {Bhagwat}, S. and {Bhandare}, R. and {Bilenko}, I.~A. and {Billingsley}, G. and {Birch}, J. and {Birney}, I.~A. and {Birnholtz}, O. and {Biscans}, S. and {Bisht}, A. and {Bitossi}, M. and {Biwer}, C. and {Bizouard}, M.~A. and {Blackburn}, J.~K. and {Blair}, C.~D. and {Blair}, D.~G. and {Blair}, R.~M. and {Bloemen}, S. and {Bock}, O. and {Bodiya}, T.~P. and {Boer}, M. and {Bogaert}, G. and {Bogan}, C. and {Bohe}, A. and {Bojtos}, P. and {Bond}, C. and {Bondu}, F. and {Bonnand}, R. and {Boom}, B.~A. and {Bork}, R. and {Boschi}, V. and {Bose}, S. and {Bouffanais}, Y. and {Bozzi}, A. and {Bradaschia}, C. and {Brady}, P.~R. and {Braginsky}, V.~B. and {Branchesi}, M. and {Brau}, J.~E. and {Briant}, T. and {Brillet}, A. and {Brinkmann}, M. and {Brisson}, V. and {Brockill}, P. and {Brooks}, A.~F. and {Brown}, D.~A. and {Brown}, D.~D. and {Brown}, N.~M. and {Buchanan}, C.~C. and {Buikema}, A. and {Bulik}, T. and {Bulten}, H.~J. and {Buonanno}, A. and {Buskulic}, D. and {Buy}, C. and {Byer}, R.~L. and {Cabero}, M. and {Cadonati}, L. and {Cagnoli}, G. and {Cahillane}, C. and {Bustillo}, J. Calder{\'o}n and {Callister}, T. and {Calloni}, E. and {Camp}, J.~B. and {Cannon}, K.~C. and {Cao}, J. and {Capano}, C.~D. and {Capocasa}, E. and {Carbognani}, F. and {Caride}, S. and {Diaz}, J. Casanueva and {Casentini}, C. and {Caudill}, S. and {Cavagli{\`a}}, M. and {Cavalier}, F. and {Cavalieri}, R. and {Cella}, G. and {Cepeda}, C.~B. and {Baiardi}, L. Cerboni and {Cerretani}, G. and {Cesarini}, E. and {Chakraborty}, R. and {Chalermsongsak}, T. and {Chamberlin}, S.~J. and {Chan}, M. and {Chao}, S. and {Charlton}, P. and {Chassande-Mottin}, E. and {Chen}, H.~Y. and {Chen}, Y. and {Cheng}, C. and {Chincarini}, A. and {Chiummo}, A. and {Cho}, H.~S. and {Cho}, M. and {Chow}, J.~H. and {Christensen}, N. and {Chu}, Q. and {Chua}, S. and {Chung}, S. and {Ciani}, G. and {Clara}, F. and {Clark}, J.~A. and {Cleva}, F. and {Coccia}, E. and {Cohadon}, P. -F. and {Colla}, A. and {Collette}, C.~G. and {Cominsky}, L. and {Constancio}, M. and {Conte}, A. and {Conti}, L. and {Cook}, D. and {Corbitt}, T.~R. and {Cornish}, N. and {Corsi}, A. and {Cortese}, S. and {Costa}, C.~A. and {Coughlin}, M.~W. and {Coughlin}, S.~B. and {Coulon}, J. -P. and {Countryman}, S.~T. and {Couvares}, P. and {Cowan}, E.~E. and {Coward}, D.~M. and {Cowart}, M.~J.},
        title = "{Observation of Gravitational Waves from a Binary Black Hole Merger}",
      journal = {\prl},
         year = 2016,
        month = feb,
       volume = {116},
       number = {6},
          eid = {061102},
        pages = {061102},
          doi = {10.1103/PhysRevLett.116.061102},
archivePrefix = {arXiv},
       eprint = {1602.03837},
 primaryClass = {gr-qc},
       adsurl = {https://ui.adsabs.harvard.edu/abs/2016PhRvL.116f1102A}
}

@ARTICLE{2019A&ARv..27....4P,
       author = {{Petroff}, E. and {Hessels}, J.~W.~T. and {Lorimer}, D.~R.},
        title = "{Fast radio bursts}",
      journal = {\aapr},
         year = 2019,
        month = dec,
       volume = {27},
       number = {1},
          eid = {4},
        pages = {4},
          doi = {10.1007/s00159-019-0116-6},
archivePrefix = {arXiv},
       eprint = {1904.07947},
 primaryClass = {astro-ph.HE},
       adsurl = {https://ui.adsabs.harvard.edu/abs/2019A&ARv..27....4P}
}

@INPROCEEDINGS{2020SPIE11445E..7MY,
       author = {{Yuan}, Xiangyan and {Li}, Zhengyang and {Liu}, Xiaowei and {Niu}, Dongsheng and {Lu}, Qishui and {Jiang}, Fanghua and {Wang}, Yuefei and {Li}, Xiaoyan and {Liang}, YongJun and {Wang}, Hai and {Zhang}, Chao and {Wang}, Jinfeng and {Li}, Bo and {Tian}, Jie and {Lu}, Haiping and {Chen}, Bingqiu and {Huang}, Yang and {Liu}, Xiangkun and {Yao}, Zhengqiu and {Cui}, Xiangqun and {Li}, Guoping},
        title = "{Development of the Multi-channel Photometric Survey telescope}",
    booktitle = {Ground-based and Airborne Telescopes VIII},
         year = 2020,
       editor = {{Marshall}, Heather K. and {Spyromilio}, Jason and {Usuda}, Tomonori},
       series = {Society of Photo-Optical Instrumentation Engineers (SPIE) Conference Series},
       volume = {11445},
        month = dec,
          eid = {114457M},
        pages = {114457M},
          doi = {10.1117/12.2562334},
       adsurl = {https://ui.adsabs.harvard.edu/abs/2020SPIE11445E..7MY}
}

@ARTICLE{2016arXiv161205560C,
       author = {{Chambers}, K.~C. and {Magnier}, E.~A. and {Metcalfe}, N. and {Flewelling}, H.~A. and {Huber}, M.~E. and {Waters}, C.~Z. and {Denneau}, L. and {Draper}, P.~W. and {Farrow}, D. and {Finkbeiner}, D.~P. and {Holmberg}, C. and {Koppenhoefer}, J. and {Price}, P.~A. and {Rest}, A. and {Saglia}, R.~P. and {Schlafly}, E.~F. and {Smartt}, S.~J. and {Sweeney}, W. and {Wainscoat}, R.~J. and {Burgett}, W.~S. and {Chastel}, S. and {Grav}, T. and {Heasley}, J.~N. and {Hodapp}, K.~W. and {Jedicke}, R. and {Kaiser}, N. and {Kudritzki}, R. -P. and {Luppino}, G.~A. and {Lupton}, R.~H. and {Monet}, D.~G. and {Morgan}, J.~S. and {Onaka}, P.~M. and {Shiao}, B. and {Stubbs}, C.~W. and {Tonry}, J.~L. and {White}, R. and {Ba{\~n}ados}, E. and {Bell}, E.~F. and {Bender}, R. and {Bernard}, E.~J. and {Boegner}, M. and {Boffi}, F. and {Botticella}, M.~T. and {Calamida}, A. and {Casertano}, S. and {Chen}, W. -P. and {Chen}, X. and {Cole}, S. and {Deacon}, N. and {Frenk}, C. and {Fitzsimmons}, A. and {Gezari}, S. and {Gibbs}, V. and {Goessl}, C. and {Goggia}, T. and {Gourgue}, R. and {Goldman}, B. and {Grant}, P. and {Grebel}, E.~K. and {Hambly}, N.~C. and {Hasinger}, G. and {Heavens}, A.~F. and {Heckman}, T.~M. and {Henderson}, R. and {Henning}, T. and {Holman}, M. and {Hopp}, U. and {Ip}, W. -H. and {Isani}, S. and {Jackson}, M. and {Keyes}, C.~D. and {Koekemoer}, A.~M. and {Kotak}, R. and {Le}, D. and {Liska}, D. and {Long}, K.~S. and {Lucey}, J.~R. and {Liu}, M. and {Martin}, N.~F. and {Masci}, G. and {McLean}, B. and {Mindel}, E. and {Misra}, P. and {Morganson}, E. and {Murphy}, D.~N.~A. and {Obaika}, A. and {Narayan}, G. and {Nieto-Santisteban}, M.~A. and {Norberg}, P. and {Peacock}, J.~A. and {Pier}, E.~A. and {Postman}, M. and {Primak}, N. and {Rae}, C. and {Rai}, A. and {Riess}, A. and {Riffeser}, A. and {Rix}, H.~W. and {R{\"o}ser}, S. and {Russel}, R. and {Rutz}, L. and {Schilbach}, E. and {Schultz}, A.~S.~B. and {Scolnic}, D. and {Strolger}, L. and {Szalay}, A. and {Seitz}, S. and {Small}, E. and {Smith}, K.~W. and {Soderblom}, D.~R. and {Taylor}, P. and {Thomson}, R. and {Taylor}, A.~N. and {Thakar}, A.~R. and {Thiel}, J. and {Thilker}, D. and {Unger}, D. and {Urata}, Y. and {Valenti}, J. and {Wagner}, J. and {Walder}, T. and {Walter}, F. and {Watters}, S.~P. and {Werner}, S. and {Wood-Vasey}, W.~M. and {Wyse}, R.},
        title = "{The Pan-STARRS1 Surveys}",
      journal = {arXiv e-prints},
         year = 2016,
        month = dec,
          eid = {arXiv:1612.05560},
        pages = {arXiv:1612.05560},
          doi = {10.48550/arXiv.1612.05560},
archivePrefix = {arXiv},
       eprint = {1612.05560},
 primaryClass = {astro-ph.IM},
       adsurl = {https://ui.adsabs.harvard.edu/abs/2016arXiv161205560C}
}

@ARTICLE{2012AcASn..53..249S,
       author = {{Sun}, R.~Y. and {Ping}, Y.~D. and {Zhao}, C.~Y.},
        title = "{A New Method for Detecting GEO Space Objects with Image Stacking}",
      journal = {Acta Astronomica Sinica},
         year = 2012,
        month = may,
       volume = {53},
       number = {3},
        pages = {249-258},
       adsurl = {https://ui.adsabs.harvard.edu/abs/2012AcASn..53..249S}
}

@ARTICLE{2005PABei..23..304M,
       author = {{Mao}, Yin-Dun and {Tang}, Zheng-Hong and {Zheng}, Yi-Jin and {Cao}, Kai},
        title = "{The basic principle and the application in astronomy of CCD drift-scan}",
      journal = {Progress in Astronomy},
         year = 2005,
        month = dec,
       volume = {23},
       number = {4},
        pages = {304-317},
       adsurl = {https://ui.adsabs.harvard.edu/abs/2005PABei..23..304M}
}

@ARTICLE{2024AJ....168...94S,
       author = {{Suggs}, Christiana Z. and {Hintz}, Eric G. and {Stephens}, Denise C.},
        title = "{Establishing Baselines for Six Short-period {\ensuremath{\delta}} Scuti Variables}",
      journal = {\aj},
         year = 2024,
        month = aug,
       volume = {168},
       number = {2},
          eid = {94},
        pages = {94},
          doi = {10.3847/1538-3881/ad5a81},
       adsurl = {https://ui.adsabs.harvard.edu/abs/2024AJ....168...94S}
}

@INPROCEEDINGS{2008SPIE.7012E..2DC,
       author = {{Cui}, Xiangqun and {Yuan}, Xiangyan and {Gong}, Xuefei},
        title = "{Antarctic Schmidt Telescopes (AST3) for Dome A}",
    booktitle = {Ground-based and Airborne Telescopes II},
         year = 2008,
       editor = {{Stepp}, Larry M. and {Gilmozzi}, Roberto},
       series = {Society of Photo-Optical Instrumentation Engineers (SPIE) Conference Series},
       volume = {7012},
        month = jul,
          eid = {70122D},
        pages = {70122D},
          doi = {10.1117/12.789458},
       adsurl = {https://ui.adsabs.harvard.edu/abs/2008SPIE.7012E..2DC}
}

@ARTICLE{2023MNRAS.520.5635Y,
       author = {{Yang}, Xu and {Hu}, Yi and {Shang}, Zhaohui and {Ma}, Bin and {Ashley}, Michael C.~B. and {Cui}, Xiangqun and {Du}, Fujia and {Fu}, Jianning and {Gong}, Xuefei and {Gu}, Bozhong and {Jiang}, Peng and {Li}, Xiaoyan and {Li}, Zhengyang and {Tao}, Charling and {Wang}, Lifan and {Xu}, Lingzhe and {Yang}, Shi-hai and {Yu}, Ce and {Yuan}, Xiangyan and {Zhou}, Ji-lin and {Zhu}, Zhenxi},
        title = "{Data release of the AST3-2 automatic survey from Dome A, Antarctica}",
      journal = {\mnras},
         year = 2023,
        month = apr,
       volume = {520},
       number = {4},
        pages = {5635-5650},
          doi = {10.1093/mnras/stad498},
archivePrefix = {arXiv},
       eprint = {2302.06997},
 primaryClass = {astro-ph.SR},
       adsurl = {https://ui.adsabs.harvard.edu/abs/2023MNRAS.520.5635Y}
}

@ARTICLE{2023SCPMA..6609512W,
       author = {{Wang}, Tinggui and {Liu}, Guilin and {Cai}, Zhenyi and {Geng}, Jinjun and {Fang}, Min and {He}, Haoning and {Jiang}, Ji-an and {Jiang}, Ning and {Kong}, Xu and {Li}, Bin and {Li}, Ye and {Luo}, Wentao and {Pan}, Zhizheng and {Wu}, Xuefeng and {Yang}, Ji and {Yu}, Jiming and {Zheng}, Xianzhong and {Zhu}, Qingfeng and {Cai}, Yi-Fu and {Chen}, Yuanyuan and {Chen}, Zhiwei and {Dai}, Zigao and {Fan}, Lulu and {Fan}, Yizhong and {Fang}, Wenjuan and {He}, Zhicheng and {Hu}, Lei and {Hu}, Maokai and {Jin}, Zhiping and {Jiang}, Zhibo and {Li}, Guoliang and {Li}, Fan and {Li}, Xuzhi and {Liang}, Runduo and {Lin}, Zheyu and {Liu}, Qingzhong and {Liu}, Wenhao and {Liu}, Zhengyan and {Liu}, Wei and {Liu}, Yao and {Lou}, Zheng and {Qu}, Han and {Sheng}, Zhenfeng and {Shi}, Jianchun and {Shu}, Yiping and {Su}, Zhenbo and {Sun}, Tianrui and {Wang}, Hongchi and {Wang}, Huiyuan and {Wang}, Jian and {Wang}, Junxian and {Wei}, Daming and {Wei}, Junjie and {Xue}, Yongquan and {Yan}, Jingzhi and {Yang}, Chao and {Yuan}, Ye and {Yuan}, Yefei and {Zhang}, Hongxin and {Zhang}, Miaomiao and {Zhao}, Haibin and {Zhao}, Wen},
        title = "{Science with the 2.5-meter Wide Field Survey Telescope (WFST)}",
      journal = {Science China Physics, Mechanics, and Astronomy},
         year = 2023,
        month = oct,
       volume = {66},
       number = {10},
          eid = {109512},
        pages = {109512},
          doi = {10.1007/s11433-023-2197-5},
archivePrefix = {arXiv},
       eprint = {2306.07590},
 primaryClass = {astro-ph.IM},
       adsurl = {https://ui.adsabs.harvard.edu/abs/2023SCPMA..6609512W}
}

@INPROCEEDINGS{2019gage.confE..14L,
       author = {{Liu}, X.},
       title ="{Multi-channel Photometric Survey Telescope-Mephisto}",
    booktitle = {Galactic Archaeology in the Gaia Era},
         year = 2019,
        month = jan,
          eid = {14},
        pages = {14},
       adsurl = {https://ui.adsabs.harvard.edu/abs/2019gage.confE..14L}
}

@ARTICLE{2017SciBu..62.1433H,
       author = {{Hu}, Lei and {Wu}, Xuefeng and {Andreoni}, Igor and {Ashley}, Michael C.~B. and {Cooke}, Jeff and {Cui}, Xiangqun and {Du}, Fujia and {Dai}, Zigao and {Gu}, Bozhong and {Hu}, Yi and {Lu}, Haiping and {Li}, Xiaoyan and {Li}, Zhengyang and {Liang}, Ensi and {Liu}, Liangduan and {Ma}, Bin and {Shang}, Zhaohui and {Sun}, Tianrui and {Suntzeff}, N.~B. and {Tao}, Charling and {Udden}, Syed A. and {Wang}, Lifan and {Wang}, Xiaofeng and {Wen}, Haikun and {Xiao}, Di and {Su}, Jin and {Yang}, Ji and {Yang}, Shihai and {Yuan}, Xiangyan and {Zhou}, Hongyan and {Zhang}, Hui and {Zhou}, Jilin and {Zhu}, Zonghong},
        title = "{Optical observations of LIGO source GW 170817 by the Antarctic Survey Telescopes at Dome A, Antarctica}",
      journal = {Science Bulletin},
         year = 2017,
        month = oct,
       volume = {62},
        pages = {1433-1438},
          doi = {10.1016/j.scib.2017.10.006},
archivePrefix = {arXiv},
       eprint = {1710.05462},
 primaryClass = {astro-ph.HE},
       adsurl = {https://ui.adsabs.harvard.edu/abs/2017SciBu..62.1433H}
}

@ARTICLE{2022RAA....22b5004L,
       author = {{Lei}, Lei and {Chen}, Bing-Qiu and {Li}, Jin-Da and {Wu}, Jin-Tai and {Jiang}, Si-Yi and {Liu}, Xiao-Wei},
        title = "{Identifications of RR Lyrae Stars and Quasars from the Simulated Data of Mephisto-W Survey}",
      journal = {Research in Astronomy and Astrophysics},
         year = 2022,
        month = feb,
       volume = {22},
       number = {2},
          eid = {025004},
        pages = {025004},
          doi = {10.1088/1674-4527/ac3adc},
archivePrefix = {arXiv},
       eprint = {2111.08316},
 primaryClass = {astro-ph.IM},
       adsurl = {https://ui.adsabs.harvard.edu/abs/2022RAA....22b5004L}
}

@ARTICLE{2025ApJS..278...29L,
       author = {{Lin}, Jie and {Wang}, Tinggui and {Cai}, Minxuan and {Wan}, Zhen and {Li}, Xuzhi and {Fan}, Lulu and {Zhu}, Qingfeng and {Jiang}, Ji-an and {Jiang}, Ning and {Kong}, Xu and {Lin}, Zheyu and {Zhu}, Jiazheng and {Liu}, Zhengyan and {Gao}, Jie and {Li}, Bin and {Li}, Feng and {Liang}, Ming and {Liu}, Hao and {Liu}, Wei and {Luo}, Wentao and {Tang}, Jinlong and {Wang}, Hairen and {Wang}, Jian and {Xue}, Yongquan and {Yao}, Dazhi and {Zhang}, Hongfei and {Zhang}, Xiaoling and {Zhao}, Wen and {Zheng}, Xianzhong},
        title = "{Minute-cadence Observations of the Galactic Plane with the Wide Field Survey Telescope (WFST): Overview, Methodology, and Early Results}",
      journal = {\apjs},
         year = 2025,
        month = may,
       volume = {278},
       number = {1},
          eid = {29},
        pages = {29},
          doi = {10.3847/1538-4365/adc0fc},
archivePrefix = {arXiv},
       eprint = {2412.12601},
 primaryClass = {astro-ph.SR},
       adsurl = {https://ui.adsabs.harvard.edu/abs/2025ApJS..278...29L}
}

@ARTICLE{2024RAA....24a5003C,
       author = {{Chen}, Yan-Peng and {Jiang}, Ji-An and {Luo}, Wen-Tao and {Zheng}, Xian-Zhong and {Fang}, Min and {Yang}, Chao and {Hong}, Yuan-Yu and {L{\"u}}, Zong-Fei},
        title = "{Basic Survey Scheduling for the Wide Field Survey Telescope (WFST)}",
      journal = {Research in Astronomy and Astrophysics},
         year = 2024,
        month = jan,
       volume = {24},
       number = {1},
          eid = {015003},
        pages = {015003},
          doi = {10.1088/1674-4527/ad07cd},
archivePrefix = {arXiv},
       eprint = {2312.03421},
 primaryClass = {astro-ph.IM},
       adsurl = {https://ui.adsabs.harvard.edu/abs/2024RAA....24a5003C}
}

@ARTICLE{2008SerAJ.176....1I,
       author = {{Ivezic}, Z. and {Axelrod}, T. and {Brandt}, W.~N. and {Burke}, D.~L. and {Claver}, C.~F. and {Connolly}, A. and {Cook}, K.~H. and {Gee}, P. and {Gilmore}, D.~K. and {Jacoby}, S.~H. and {Jones}, R.~L. and {Kahn}, S.~M. and {Kantor}, J.~P. and {Krabbendam}, V.~V. and {Lupton}, R.~H. and {Monet}, D.~G. and {Pinto}, P.~A. and {Saha}, A. and {Schalk}, T.~L. and {Schneider}, D.~P. and {Strauss}, M.~A. and {Stubbs}, C.~W. and {Sweeney}, D. and {Szalay}, A. and {Thaler}, J.~J. and {Tyson}, J.~A. and {LSST Collaboration}},
        title = "{Large Synoptic Survey Telescope: From Science Drivers To Reference Design}",
      journal = {Serbian Astronomical Journal},
         year = 2008,
        month = jun,
       volume = {176},
        pages = {1-13},
          doi = {10.2298/SAJ0876001I},
       adsurl = {https://ui.adsabs.harvard.edu/abs/2008SerAJ.176....1I}
}

@ARTICLE{2019PASP..131f8003B,
       author = {{Bellm}, Eric C. and {Kulkarni}, Shrinivas R. and {Barlow}, Tom and {Feindt}, Ulrich and {Graham}, Matthew J. and {Goobar}, Ariel and {Kupfer}, Thomas and {Ngeow}, Chow-Choong and {Nugent}, Peter and {Ofek}, Eran and {Prince}, Thomas A. and {Riddle}, Reed and {Walters}, Richard and {Ye}, Quan-Zhi},
        title = "{The Zwicky Transient Facility: Surveys and Scheduler}",
      journal = {\pasp},
         year = 2019,
        month = jun,
       volume = {131},
       number = {1000},
        pages = {068003},
          doi = {10.1088/1538-3873/ab0c2a},
archivePrefix = {arXiv},
       eprint = {1905.02209},
 primaryClass = {astro-ph.IM},
       adsurl = {https://ui.adsabs.harvard.edu/abs/2019PASP..131f8003B}
}

@ARTICLE{2019PASP..131a8002B,
       author = {{Bellm}, Eric C. and {Kulkarni}, Shrinivas R. and {Graham}, Matthew J. and {Dekany}, Richard and {Smith}, Roger M. and {Riddle}, Reed and {Masci}, Frank J. and {Helou}, George and {Prince}, Thomas A. and {Adams}, Scott M. and {Barbarino}, C. and {Barlow}, Tom and {Bauer}, James and {Beck}, Ron and {Belicki}, Justin and {Biswas}, Rahul and {Blagorodnova}, Nadejda and {Bodewits}, Dennis and {Bolin}, Bryce and {Brinnel}, Valery and {Brooke}, Tim and {Bue}, Brian and {Bulla}, Mattia and {Burruss}, Rick and {Cenko}, S. Bradley and {Chang}, Chan-Kao and {Connolly}, Andrew and {Coughlin}, Michael and {Cromer}, John and {Cunningham}, Virginia and {De}, Kishalay and {Delacroix}, Alex and {Desai}, Vandana and {Duev}, Dmitry A. and {Eadie}, Gwendolyn and {Farnham}, Tony L. and {Feeney}, Michael and {Feindt}, Ulrich and {Flynn}, David and {Franckowiak}, Anna and {Frederick}, S. and {Fremling}, C. and {Gal-Yam}, Avishay and {Gezari}, Suvi and {Giomi}, Matteo and {Goldstein}, Daniel A. and {Golkhou}, V. Zach and {Goobar}, Ariel and {Groom}, Steven and {Hacopians}, Eugean and {Hale}, David and {Henning}, John and {Ho}, Anna Y.~Q. and {Hover}, David and {Howell}, Justin and {Hung}, Tiara and {Huppenkothen}, Daniela and {Imel}, David and {Ip}, Wing-Huen and {Ivezi{\'c}}, {\v{Z}}eljko and {Jackson}, Edward and {Jones}, Lynne and {Juric}, Mario and {Kasliwal}, Mansi M. and {Kaspi}, S. and {Kaye}, Stephen and {Kelley}, Michael S.~P. and {Kowalski}, Marek and {Kramer}, Emily and {Kupfer}, Thomas and {Landry}, Walter and {Laher}, Russ R. and {Lee}, Chien-De and {Lin}, Hsing Wen and {Lin}, Zhong-Yi and {Lunnan}, Ragnhild and {Giomi}, Matteo and {Mahabal}, Ashish and {Mao}, Peter and {Miller}, Adam A. and {Monkewitz}, Serge and {Murphy}, Patrick and {Ngeow}, Chow-Choong and {Nordin}, Jakob and {Nugent}, Peter and {Ofek}, Eran and {Patterson}, Maria T. and {Penprase}, Bryan and {Porter}, Michael and {Rauch}, Ludwig and {Rebbapragada}, Umaa and {Reiley}, Dan and {Rigault}, Mickael and {Rodriguez}, Hector and {van Roestel}, Jan and {Rusholme}, Ben and {van Santen}, Jakob and {Schulze}, S. and {Shupe}, David L. and {Singer}, Leo P. and {Soumagnac}, Maayane T. and {Stein}, Robert and {Surace}, Jason and {Sollerman}, Jesper and {Szkody}, Paula and {Taddia}, F. and {Terek}, Scott and {Van Sistine}, Angela and {van Velzen}, Sjoert and {Vestrand}, W. Thomas and {Walters}, Richard and {Ward}, Charlotte and {Ye}, Quan-Zhi and {Yu}, Po-Chieh and {Yan}, Lin and {Zolkower}, Jeffry},
        title = "{The Zwicky Transient Facility: System Overview, Performance, and First Results}",
      journal = {\pasp},
         year = 2019,
        month = jan,
       volume = {131},
       number = {995},
        pages = {018002},
          doi = {10.1088/1538-3873/aaecbe},
archivePrefix = {arXiv},
       eprint = {1902.01932},
 primaryClass = {astro-ph.IM},
       adsurl = {https://ui.adsabs.harvard.edu/abs/2019PASP..131a8002B}
}

@ARTICLE{2018MNRAS.479..111M,
       author = {{Ma}, Bin and {Shang}, Zhaohui and {Hu}, Yi and {Hu}, Keliang and {Liu}, Qiang and {Ashley}, Michael C.~B. and {Cui}, Xiangqun and {Du}, Fujia and {Fan}, Dongwei and {Feng}, Longlong and {Huang}, Fang and {Gu}, Bozhong and {He}, Boliang and {Ji}, Tuo and {Li}, Xiaoyan and {Li}, Zhengyang and {Liu}, Huigen and {Tian}, Qiguo and {Tao}, Charling and {Wang}, Daxing and {Wang}, Lifan and {Wang}, Songhu and {Wang}, Xiaofeng and {Wei}, Peng and {Wu}, Jianghua and {Xu}, Lingzhe and {Yang}, Shihai and {Yang}, Ming and {Yang}, Yi and {Yu}, Ce and {Yuan}, Xiangyan and {Zhou}, Hongyan and {Zhang}, Hui and {Zhang}, Xueguang and {Zhang}, Yi and {Zhao}, Cheng and {Zhou}, Jilin and {Zhu}, Zong-Hong},
        title = "{The first release of the AST3-1 Point Source Catalogue from Dome A, Antarctica}",
      journal = {\mnras},
         year = 2018,
        month = sep,
       volume = {479},
       number = {1},
        pages = {111-120},
          doi = {10.1093/mnras/sty1392},
archivePrefix = {arXiv},
       eprint = {1805.05566},
 primaryClass = {astro-ph.IM},
       adsurl = {https://ui.adsabs.harvard.edu/abs/2018MNRAS.479..111M}
}

@ARTICLE{2020RAA....20..168S,
       author = {{Shang}, Zhaohui},
        title = "{Astronomy from Dome A in Antarctica}",
      journal = {Research in Astronomy and Astrophysics},
         year = 2020,
        month = oct,
       volume = {20},
       number = {10},
          eid = {168},
        pages = {168},
          doi = {10.1088/1674-4527/20/10/168},
archivePrefix = {arXiv},
       eprint = {2010.04972},
 primaryClass = {astro-ph.IM},
       adsurl = {https://ui.adsabs.harvard.edu/abs/2020RAA....20..168S}
}
\bibliographystyle{aasjournal}



\end{document}